\documentclass[fleqn,useAMS,usenatbib]{mn2e}

\usepackage{times}  % changed from Times ADH

\usepackage{natbib}

\usepackage{amsfonts}
\usepackage{graphicx}
\usepackage{amssymb}
\usepackage{amsmath}
\usepackage[T1]{fontenc}

\def\const{{\rm const}}

\newcommand{\ar}{{\acute r}}
\newcommand{\al}{{\acute l}}

\newcommand{\as}{{\acute s}}
\newcommand{\am}{{\acute m}}
\newcommand{\oam}{{\acute{\overline m}}}

\newcommand{\news}{{N}}
\newcommand{\mnews}{{|N|_{\rm av}}}
\newcommand{\rr}{{\mathfrak r}}

\title[Light, gravity and pulse-time offsets]{Light rays, gravitational waves, and pulse-time offsets}
\author[Adam D. Helfer]{Adam D. Helfer\thanks{E-mail:
helfera@missouri.edu} \\
Department of Mathematics, University of Missouri, Columbia, MO 65211 U.S.A.
}
\begin{document}

\date{Accepted xxxx. Received xxxx; in original form xxxx}

\pagerange{\pageref{firstpage}--\pageref{lastpage}} \pubyear{2012}

\maketitle

\label{firstpage}

\begin{abstract}
One might expect light to be scattered when it passes through a gravitational wave, and might hope that in favourable circumstances these scatterings could be observed on Earth even if the interaction occurs far away.  
Damour and Esposito-Far\`ese, and Kopeikin, Sch\"afer, Gwinn and Eubanks, found that there were cancellations making such effects disappointingly small.
Here I show that those cancellations depend on the emission of the light occurring far behind the gravity-wave source; for light-emissions near that source, 
larger effects are possible. 
I first develop a covariant treatment of the problem in exact general relativity (the propagation of light being modelled by geometric optics), and then specialise to linearised gravity.
The most promising candidates identified here for detection in the not-too-distant future would involve sufficiently tight binaries as sources of gravitational radiation, and nearby pulsars as light-sources.  In some favourable but not extreme cases, I find offsets in the pulses' times of arrival at Earth by $\sim 10^{-10}-10^{-9}\, {\rm s}$, with periods half the binaries' periods.
\end{abstract}


\section{Introduction}

The detection of gravitational waves is a critical test of Einstein's theory of general relativity.  We have at present good indirect evidence for the existence of such waves in the orbital decays of binary pulsars~\citep{TaylorWeissberg}, but no direct evidence.  Unfortunately, we expect most astrophysical sources to be so distant that the waves should be very weak by the time they reach the solar system, making terrestrial detection a challenge.  

It is natural to ask whether gravitational waves might have effects on matter or light closer to the sources, where the waves are stronger, but with the effects being observable on the Earth.  In particular, one might search for modulations in light caused by its passage through gravitational waves, for instance light from
another electromagnetic source passing nearby a neutron star or binary system.  (Here and throughout, `light' means electromagnetic signals of any frequency which can be treated by geometric optics.)
These modulations
could include variations in intensity, in apparent position, in polarisation, in phase, or in time-dilation (often referred to in the literature as time-delay).  Of course, such variations could also be produced by other means; one would need further information to conclude they were caused by gravitational waves.  

\subsection{Previous work and its implications}

Such possibilities were contemplated by a number of authors, notably \citet{Sazhin1978}, \citet{Lab1993}, \citet{Dur1994} and \citet{Fak1994}; it seemed from these papers that some of the effects might be detectable in the near future.   Concerns, though, were raised about the adequacy of those treatments, and \citet{DEF1998} and \citet*{KSGU} (see also \citealt{KJ1997}), after more considered analyses, 
found very much smaller effects than then the optimistic ones in the earlier papers.\footnote{The literature on this subject is extensive; for more references and discussion, see \citet{KJ1997,DEF1998,KSGU}.    
Additional references, on the light's polarisation, are \citet{KM2002,PM2002,Faraoni2008}; and \citet{LLMM} on the light's phase.} 
The effects were so small that they seemed to preclude the detection of gravitational waves by these means, at least for known classes of systems in the near future.
This conclusion has been
accepted by much of the community \citep{Schutz2010,BF2011}.

But the situation is not so clear-cut.  For one thing, the later authors' investigations, while overlapping the earlier ones, do not cover all of the physical possibilities they considered.  Also the results as they stand strongly suggest there are essential points which have not yet been understood.

We can appreciate this if we ask how such gross discrepancies between the earlier and later work could have occurred.  While it is true that much of the earlier work involved rough approximations, one would think that in order to explain such a large and systematic change there must have been a shift in what was taken to be the essential physics.

The earlier papers typically and very plausibly found that the light rays responded to the gravitational radiation fields of the sources, but later authors found that there were cancellations which resulted in those terms not entering.\footnote{It is worth noting that essentially the same feature comes up in the textbook analysis of the geodesic deviation of linearised plane waves in a standard gauge, used to discuss gravitational wave detectors, where one finds that the deviation depends on the metric coefficients at the end-points only of the geodesic~\citep{MTW}. }
Indeed, Damour and Esposito-Far\`ese pointed out that pure, vacuum, gravitational waves would give {\em no} overall scattering in linearized theory.
More specifically, both Damour and Esposito-Far\`ese, and Kopeikin et al., reported that what scattering of a light-ray by an isolated source (on a straight world-line in linearised theory) does occur depends only on the source's configuration at the light-ray's point of closest approach.  This is something which cannot be locally causally determined, and therefore must depend on a non-local cancellation of effects over the light-ray's trajectory.

These observations suggest that the cancellations are tied to the infinite extent of the light-ray's trajectory, and may not occur for propagation along finite, or only half-infinite, segments.\footnote{Since `light-ray' has an established meaning as a path, whether finite, half-infinite or doubly-infinite, taken by light in geometric optics, I will not use `ray' to mean a half-infinite segment.}  We shall see below that this is indeed the case.  
In other words, the cancellations are present when the light effectively comes in from infinity, but larger effects are possible when the emission is from a finite point, in particular when it might be close to the gravitational-wave source.
Such situations are very much of interest, because there are many astrophysical configurations in which there is a source of light near an expected source of gravitational radiation.

This fits well with other points in the literature.
One can see that many of the earlier works did, either explicitly or in effect, use scattering estimates which depended on integration over half-infinite or finite segments rather than doubly-infinite lines, and in some cases (depending on the situation considered) the authors were correct in wanting to do so.  And Damour and Esposito-Far\`ese (see also \citealt{Faraoni2008}) correctly noted that overcoming the cancellations would depend on `edge effects'.

The paper of Kopeikin et al. is sometimes considered to show that no significantly larger effects arise from finite or half-infinite segments.  This, however, is not really correct.  While these authors did indeed work out formulas for the null geodesics around certain sources in linearised gravity, those formulas are very lengthy and were examined in detail only in certain limiting cases.  None of those cases corresponds to the configuration of ultimate interest here, a light-emission reasonably close to the gravitational-wave source (precisely, within of the order of the impact parameter to the light-ray's point of closest approach to the gravity-wave source) received by a distant observer.  
In fact, the computations of Kopeikin et al. are consistent with the possibility of larger effects from restricted segments.

There is, additionally, a key issue which must be properly addressed if we are to take into account the emission (or reception) of light-rays at finite events, especially events near gravitationally radiating sources.  We do not expect to learn anything about gravitational radiation from the reception of one 
light-ray; rather, it is in the changes in the light received over time
that the we hope the information is encoded.  In other words, it is {\em differential} scattering effects, of 
successive light-rays {\em relative} to each other,
which are of interest.  But to treat these properly, we must know we are modelling the emissions of the successive rays accurately.  This is potentially problematic when the light-sources themselves are in the region affected by the gravitational waves, because the effects of the waves on the light through their perturbations of the motion of the light-sources compete with the other contributions.  

Another way of saying this is that the received signal depends, not only on the gravitational field through which the light has passed, but also on the apparent acceleration of the light-emitter, as inferred from electromagnetic measurements at the receiver.  Even if the emitter suffers no {\em local} acceleration, its trajectory as reconstructed from light-signals coming into the detector will appear to involve accelerations, because of the gravitational radiation the light has passed through.  (And so one might want to say that the variations in the received signal are not purely due to scattering, but also to this non-local acceleration of the source.  However, from a general-relativistic point of view these are two facets of the same thing.)

This is a point which has only been partially attended to in the literature on scattering.\footnote{Curiously, there is another line of thought on gravitational-wave detection, where the proper resolution of this point is an integral, if little-emphasized, part of the foundations.  This is the idea of pulsar timing arrays.  Often the basic principle is explained as the use of the pulsar and the Earth as the two ends of a gravitational-wave detector:  thus the emissions of pulses are from events along the pulsar's world-line, which is freely falling (e.g. \citealt{HD1983}).}  
In certain restricted cases, where the gravitational waves have a simple form and the light-rays considered are aligned to take advantage of it, the issue has been considered (e.g.~\citealt{Sazhin1978}), but the
pattern 
has been to compute the null geodesics but not examine all the details of the time-dependence of light-emission and reception.  

For instance, \citet{KSGU} worked out the null geodesics around certain gravitational sources in linearised theory, in terms of the geodesics' initial positions and tangents.
They derived formulas for the coordinates of finite points along the geodesics, but they did not fully resolve the question of how to choose the data for a one-parameter family of null geodesics emitted at one world-line and received at another. 
(They addressed some aspects of the issue, with their discussion of the appropriateness of their gauge choice.  That argument shows that some light-emitters may be modelled by taking the spatial coordinates, in the gauge used, fixed.  For such an emitter, the initial positions of the null geodesics would simply be the points on the emitter's world-line.  However, this leaves open the question of how to choose the geodesics' tangents, so as to ensure the geodesics go from the emitter to the receiver; it also leaves open the question of how to treat generic emitters, which need not have their space coordinates constant.)
Developments of this approach can of course take into such variations; see \citet{KEK2011}.

We wish then to extend the existing treatment of light-scattering by gravitational fields to propagation over finite or half-infinite segments, accounting for the gravitational waves' effects on the light-sources in those cases.  It is worthwhile considering what the best approach is.

\subsection{A more general setting}

Virtually all work that has been done in this area has been done in the linearised-gravity (or at least linearised gravitational-wave) setting, and has used detailed coordinate calculations in specific gauges.
There are reasons for wanting a more general, and invariant, approach.

Most importantly, the core issue -- how light emitted from successive events along one world-line is received along another, after passage through a region of space--time -- is evidently a basic problem in relativity:  it is worth treating in generality. It is a tenet of relativity that the physics is invariant, so invariant treatments focus most directly on the physics; a central goal of this paper is to give one.

As emphasized earlier, the key physical observables are associated with {\em differential} scattering, of successive light-rays relative to each other.  This is governed by the geodesic deviation equation, a linear ordinary differential equation involving the curvature, and it is the solution of this equation which is the central mathematical chore in an invariant approach.  By contrast, gauge-based approaches focus on the metric and on integrating the null geodesics, a considerable task.  It is only once these are known that one can proceed to
extract the measurable differential effects (and, as noted above, this further step has not generally been completely worked out in these approaches).
One sees that the invariant approach goes more directly to the observables.  Also, by limiting oneself to differential effects, one expands considerably the cases which can be treated, since one need not confront the problem of finding the null geodesics explicitly.
The invariant formulas derived here would, for instance, be applicable in a cosmological context, or to treat light-rays in strong gravitational fields.

With the invariant formulas in hand, one can always specialise to gauge choices which may be convenient for particular problems.  But not having to make such a choice too early saves much work, and does away with many conceptual issues associated with gauges.

In some of the literature the motivation for the choice of gauge is to be confident that one is indeed modelling gravitational waves, as opposed to other gravitational disturbances.  In this paper, that is a distinction which is best made, not at the beginning of the analysis, but rather once we have the general formulas.
This is because there is no very simple invariant characterization of what counts as gravitational radiation and what does not.  We first derive general, invariant, formulas; we may then specialise these to systems which are gravitationally radiating.

\subsection{Goals and results}

The aims and results of this paper are:

(a) To provide an invariant treatment of the effects of gravity on light, in the approximation where the electromagnetic field can be treated by geometric optics, in exact general relativity, and a specialisation of that treatment to linearised gravity.  In particular, while much of the motivation for this work comes from wanting to understand gravitational waves, the approach here allows us to treat any gravitational fields in general relativity.

I emphasized above that the effects of interest were all differential ones, where one light-ray is compared with a nearby one.  Thus the main tool in the analysis is the geodesic deviation equation.  Essentially, the task is to solve this equation with initial and final conditions corresponding to the emission of light at one world-line and its absorption at another.

This approach allows us to disregard many of the coordinate issues which occur elsewhere.  For instance, previous papers have tended to discuss `time-delay' effects in terms of local coordinate times, whereas it is really the clock times of the emitter and the receiver of light which are relevant; the approach here gives an integro-differential equation directly relating the clock times.  

There are similar equations for the changes observed at the receiver in the wave-vector (this is equivalent to equations for the change in phase, since the wave-vector is the gradient of the phase) and the electromagnetic field (and thus the amplitudes and polarisations).  In all of these cases, the changes are determined partly by the geometry of the space--time the light propagates through, partly by the world-lines of the emitter and receiver, and partly by any intrinsic changes in the field at the emitter.  All of these contributions are accounted for directly.

These results are closely related to those of Damour and Esposito-Far\`ese; the chief difference is that they worked with the field's Fourier transform and gave global results, whereas our analysis is directly in terms of the radiative data and considers not only the limiting case of a doubly-infinite geodesic, but how that limit arises.
Similarly, the results here are also compatible with the computations of Kopeikin et al.

(b) To give rough estimates of the effects in question in the case of quadrupolar radiation in linearised gravity.

Of the various effects on the propagation of light by its passage through, or emission in, a gravitational wave field -- changes in its position on the celestial sphere of the receiver, in its phase, polarisation, amplitude, or time-dilation -- the last is, at present, the least difficult to imagine measuring.  We shall find that changes in the position on the celestial sphere and phase are suppressed by a geometric factor (related to the `lighthouse effect').  While changes in polarisation or (fractional) amplitude might be, in numerical terms, of the same magnitudes as those due to time-dilation, it is hard to foresee polarisation or intensity measurements accurate enough to reveal gravitational-wave effects.

The time-dilation effects (often referred to as `time-delay effects') arise from the distortion $\tau _1(\tau )$ of the light-emitter's proper time $\tau _1$ as measured by light-signals arriving at the receiver's proper time $\tau$.  This function $\tau _1(\tau )$ depends on gravitational radiation the light encounters; in particular, we shall see that an integral of a curvature component over the geodesic contributes to $d^2\tau _1/d\tau ^2$.

One result of this is a red-shift; also, if the emitter can be expected to give off signals at known proper times (for instance, if it is a pulsar), then the light's passage through a gravitational wave may cause the times of arrival of those signals to wander slightly from what they would have been, had no radiation been present.

The detection of these effects would require training a telescope, whether optical or radio, on a light-emitter in the vicinity of a gravitational-wave source.  For this reason we are interested primarily in sources which have at least a reasonable likelihood of radiating in a reasonable observing period.  These will be continuous, localised, sources of gravitational waves, which unfortunately are thought to be weak.

The best sort of candidate found here would involve a tight binary as the source of gravitational waves, with a nearby pulsar (perhaps as a distant tertiary) as the light-emitter.  For two $5M_\odot$ stars (or black holes) in a $.1$-day circular orbit, I find possible displacements $\Delta\tau _{\rm max}$ in pulse arrival times $\sim 2\times 10^{-10}\, {\rm s}$.  For (as might possibly be found in a globular cluster) a one-solar-mass star in a one-day circular orbit around a $10^4M_\odot$ black hole, 
one has $\Delta\tau _{\rm max}\sim 3\times 10^{-9}\, {\rm s}$.  For (as might happen near the galactic centre) a star of mass $m$ in a ten-year circular orbit around a $4\times 10^6M_\odot$ black-hole, one has $\Delta\tau _{\rm max}\sim 7\times 10^{-10}(m/M_\odot)\, {\rm s}$.  These displacements would vary with a period half the orbital period.

\subsection{Organization}

The next section develops a general formalism for the analysis, 
valid in full general relativity, based on geodesic deviation.
In Section 3, this is specialised to the case of linearised gravity; there the geodesic deviation equation can be solved in terms of curvature integrals.  
Section 4 considers the effects of quadrupole sources (allowing both `electric' and `magnetic' terms) of gravitational radiation on light-rays in the sources' wave zones.  In Section 5, those results are applied to estimate the magnitudes of the effects for some expected sources of gravitational radiation.  Section 6 contains a summary and discussion.

The reader wishing to skip the derivations will find the formula (\ref{ttimeqtheta}) for the comparison of emitter and receiver clock-times at the end of section 4; the meanings of the symbols in this formula are recapitulated in the paragraph containing it.  The rough estimate of the magnitude of the effect 
is (\ref{rough}).  Sections 5 and 6, dealing with the estimates, depend (except for a few technical comments which can be skipped) only on this rough form and can be read independently of the earlier sections.

{\em Notation and conventions.}  Except where otherwise specified, the notation and conventions here are those of \citet{PR1984,PR1986}, which will also serve as a reference for all material not otherwise explained.  The metric signature is $+{}-{}-{}-$, and the curvature satisfies $[\nabla _a,\nabla _b]v^d=R_{abc}{}^dv^c$.  Factors of $c$, the speed of light, are omitted until the end of Section 4.

In keeping with this paper's main aim of an invariant treatment, the basic formalism is, in principle, the abstract-index one of Penrose and Rindler.  However, in fact, in almost all cases the tensorial indices in this paper may be interpreted equally well as abstract indices or as component indices with respect to a chart.  Where there is any difference, it is noted.

\section{General formalism}

Let us suppose we have two world-lines $\gamma _j(\tau _j)$ (with $j=1,2$) in space--time, with $\tau _j$ proper time on each.  These will be the world-lines of the source and the detector of the light-rays.\footnote{We should write $\gamma ^a_j(\tau _j)$ for the coordinates of a curve in a chart, with $a$ a coordinate index; similarly we would have $p^a(s,\tau )$ for the coordinates of a family of light-rays below.  (Since the coordinates themselves are not tensorial quantities, the quantities $\gamma _j(\tau _j)$ and $p(s,\tau )$ do not carry {\em abstract} indices.}
They need not be geodesic, and they need not be in any asymptotic or weak-field region.
Let $p(s,\tau )$ be a smooth family of light rays from $\gamma _1$ to $\gamma _2$:  for each fixed $\tau$, the light ray runs from $\gamma _1(\tau _1(\tau))=p(s_1(\tau ),\tau )$ to $\gamma _2(\tau _2(\tau ))=p(s_2(\tau ),\tau )$, with $s$ an affine parameter along the ray.  Then $l^a=\partial _sp$ will be the tangent null vector, and $w^a=\partial _\tau p$ will be the connecting, Jacobi, field.\footnote{So these relations would be $l^a=\partial _sp^a$ and $w^a=\partial _\tau p^a$ in a coordinate chart.}  
Note that we have
\begin{eqnarray}
\frac{d\tau _1}{d\tau}{\dot\gamma}_1^a&=&\frac{ds_1}{d\tau} l^a+w^a\\
\frac{d\tau _2}{d\tau} {\dot\gamma}_2^a
  &=&\frac{ds_2}{d\tau} l^a+w^a
\end{eqnarray}
on $\gamma _1$, $\gamma _2$.  We shall take $\tau=\tau _2$, so as to index the light-rays by the receiver's proper time.
There is some normalisation freedom in this:  for each $\tau$, the vector $(s_2(\tau )-s_1(\tau ))l^a$ is determined, but the individual values $s_2(\tau )$, $s_1(\tau )$, $l^a$ are not.  It will be simplest to take 
\begin{equation}
s_1(\tau )=\const\, ,\quad s_2(\tau )=\const\, ,
\end{equation}
so
\begin{equation}\label{oneq}
\frac{d\tau _1}{d\tau _2}{\dot\gamma}_1^a=w^a\mbox{ on }\gamma _1\quad\mbox{and}\quad
 {\dot\gamma}_2^a
  =w^a\mbox{ on }\gamma _2\, .
\end{equation}

We shall be interested in how the light-ray varies with $\tau$, and hence in the connecting field $w^a$.   Contracting the equation $l\cdot\nabla w^a=w\cdot\nabla l^a$ with $l^a$, we find $l\cdot\nabla (w\cdot l)=0$, which, with eq. (\ref{oneq}), gives us 
\begin{equation}\label{td}
\frac{d\tau _1}{d\tau } l\cdot {\dot\gamma}_1 = l\cdot {\dot\gamma}_2\, .
\end{equation}
It is $d\tau _1/d\tau$ which gives rise to what are often called `time-delay' effects, and eq.~(\ref{td}) will allow us to solve for these.  

We shall want the connecting vector $w^a$ in terms of the curvature along the light-rays.
For this, we must use the Jacobi equation in some detail.  Let
$U^a{}_b(s,\tau )$, $V^a{}_b(s,\tau )$ be solutions, so $(l\cdot\nabla )^2U^a{}_b=l^pl^qR_{pcq}{}^aU^c{}_b$, $(l\cdot\nabla )^2V^a{}_b=l^pl^qR_{pcq}{}^aV^c{}_b$, with 
\begin{eqnarray}
U^a{}_b(s_1,\tau )&=&\delta ^a{}_b\\ 
l\cdot\nabla U^a{}_b (s_1,\tau )&=&0\, ,
\end{eqnarray}
and
\begin{eqnarray}
V^a{}_b(s_1,\tau )&=&0\, ,\\
  l\cdot \nabla V^a{}_b(s_1,\tau )&=&\delta ^a{}_b\, .
\end{eqnarray}
Then the connecting field is
\begin{equation}\label{cveq}
w^a=U^a{}_b\alpha ^b +V^a{}_b\beta ^b\, ,
\end{equation} 
for some $\alpha ^a$, $\beta ^a$ (elements of the tangent space at $\gamma _1(\tau _1 )$).  We can find $\alpha ^a$, $\beta ^a$ by using eq. (\ref{oneq}); this gives
\begin{eqnarray}
\alpha ^a&=&\frac{d\tau _1}{d\tau}{\dot\gamma}_1^a\label{alpheq}\\
\beta ^a&=&(V(s_2)^{-1})^a{}_b\left[ {\dot\gamma}_2^b -
\frac{d\tau _1}{d\tau}
U(s_2)^b{}_c{\dot\gamma}_1^c\right]\, .\label{beteq}
\end{eqnarray}
Note that $l\cdot\nabla (l\cdot w)=0$, applied to eq. (\ref{cveq})
and evaluated at $s=s_1$, implies $l\cdot \beta =0$.

Of course, at a conjugate point $V^a{}_b$ will not be invertible; conjugate points will be discussed elsewhere.

\subsection{Time dilation}

Much of the literature is phrased in terms of `time delays', the delay being taken to be the difference in the coordinate times of emission and reception.  
This is not an invariant concept, and so it is then corrected (or corrections are at least as a matter of principle considered) to take into account differences between the clock times and the coordinate times.\footnote{The literature also refers to these as `Shapiro time delays'.  But in the Shapiro case the configuration is different:
the radio signal is reflected off a planet and then received by a detector which has (essentially) the same world-line as the emitter, and so there {\em is} a well-defined time-delay between emission and reception.  It is this difference in clock times which \citet{Shapiro1964} originally called the time delay.  The present techniques can easily be adapted to give an invariant formula for that.}

We will work with the clock times directly, the basic quantity of physical interest being $\tau _1(\tau )$, the time along $\gamma _1$ at which a signal was sent to arrive at time $\tau$ on $\gamma _2$.  Then $d\tau _1/d\tau$, the time-dilation or red-shift factor.\footnote{One has $d\tau _1/d\tau =(1+z)^{-1}$, where $z$ is the usual measure of red-shift of signals from $\gamma _1$ to $\gamma _2$.}  It is changes in this quantity which may carry the imprint of gravitational radiation.

We have
\begin{eqnarray}\label{geneq}
\frac{d^2\tau _1}{d\tau ^2}&=&w\cdot\nabla\frac{l\cdot{\dot\gamma}_2}{l\cdot{\dot\gamma}_1}\nonumber\\
&=&(l\cdot{\dot\gamma}_1)^{-2}
  \left( l\cdot{\dot\gamma}_1 w\cdot\nabla (l\cdot{\dot\gamma}_2)
    - l\cdot{\dot\gamma}_2 w\cdot\nabla (l\cdot{\dot\gamma}_1)
      \right)
     \nonumber\\ 
 &=&(l\cdot{\dot\gamma}_1)^{-2}\left(
   l\cdot{\dot\gamma}_1 {\dot\gamma}_{2a} l\cdot\nabla w^a
   -l\cdot{\dot\gamma}_2 {\dot\gamma}_{1a} l\cdot\nabla w^a\right)\nonumber\\
   &&+(l\cdot{\dot\gamma}_1)^{-1}
    l\cdot{\ddot{\gamma}_2} 
    -  (l\cdot{\dot\gamma}_1)^{-3}(l\cdot{\gamma}_2)^2
      l\cdot{\ddot{\gamma}_1}   \nonumber\\
   &=&(l\cdot{\dot\gamma}_1)^{-1}\left(
      (w_al\cdot\nabla w^a)\Bigr| _{s=s_2}
    % \right.\nonumber\\
    % &&\qquad\qquad
    %\left.
     - (w_al\cdot\nabla w^a)\Bigr| _{s=s_1}
      \right)\nonumber\\
      &&+(l\cdot{\dot\gamma}_1)^{-1}
    l\cdot{\ddot{\gamma}_2} 
    -  (l\cdot{\dot\gamma}_1)^{-3}(l\cdot{\dot\gamma}_2)^2
      l\cdot{\ddot{\gamma}_1}   \, .
\end{eqnarray}      
Expressing this in terms of $U$ and $V$, we have
\begin{eqnarray}\label{acceq}
\frac{d^2\tau _1}{d\tau ^2}&=&
(l\cdot{\dot\gamma}_1)^{-1}{\dot\gamma}_{2a}l\cdot\nabla (U^a{}_b\alpha ^b+V^a{}_b\beta ^b)\Bigr| _{s=s_2}\nonumber\\
&&-(l\cdot{\dot\gamma}_1)^{-1}(l\cdot{\dot\gamma}_2) {\dot\gamma}_1\cdot\beta\nonumber\\
&&+(l\cdot{\dot\gamma}_1)^{-1}
    l\cdot{\ddot{\gamma}_2} 
    -  (l\cdot{\dot\gamma}_1)^{-3}(l\cdot{\gamma}_2)^2
      l\cdot{\ddot{\gamma}_1}   \, .\quad
\end{eqnarray} 
The last terms of course vanish when the source and the emitter are freely falling.     
(In eqs. (\ref{geneq}), (\ref{acceq}), and subsequently, the
accelerations ${\ddot\gamma}_j^a={\dot\gamma}_j\cdot\nabla {\dot\gamma}_j^a$ are taken with respect to the proper time $\tau _j$ along the corresponding world-line.)

\subsection{Change in wave-vector}

In the geometric-optics approximation, the wave-vector is 
\begin{equation}\label{wavevec}
k^a=\omega _1l^a/l\cdot{\dot\gamma}_1\, ,
\end{equation} 
where $\omega _1$ is the angular frequency with respect to the frame of the emitter.  
We are interested in the time-dependence of $k^a$ at the detector's world-line, $\gamma _2(\tau )$.  For simplicity, we will assume here that $\omega _1$ is constant.  
(Otherwise, in what follows, one simply gets an extra term, from the product rule.)
Then applying $w\cdot\nabla$ to eq.~(\ref{wavevec}), we have
\begin{eqnarray}
{\dot\gamma}_2\cdot\nabla k^a&=&
  \omega _1(l\cdot{\dot\gamma}_1)^{-2}\left(
    l\cdot{\dot\gamma}_1 w\cdot\nabla l^a 
      -l^a w\cdot\nabla (l\cdot{\dot\gamma}_1)\right)
      \nonumber\\
  &=&\omega _1 (l\cdot {\dot\gamma}_1)^{-2}\left(
    l\cdot {\dot\gamma}_1 l\cdot\nabla w^a\Bigr| _{s_2}\right.
      \nonumber\\
      &&\quad \left. -
      l^a(w\cdot\nabla l_b){\dot\gamma}_1^b\Bigr| _{s_1}
       -l^al_bw\cdot\nabla {\dot\gamma}_1^b\Bigr| _{s_1}\right)\nonumber\\
  &=&\omega _1 (l\cdot{\dot\gamma}_1)^{-2}\left( (l\cdot{\dot\gamma}_1)
     l\cdot\nabla (U^a{}_b\alpha ^b
      +V^a{}_b\beta ^b)\Bigr| _{s_2}\right.\nonumber\\
      &&\quad\left. - l^a\left[ \beta\cdot{\dot\gamma}_1 +
      (l\cdot{\dot\gamma}_1)^{-1}
      (l\cdot{\dot\gamma}_2)  l_b {\ddot\gamma}_1^b\right]\phantom{\Bigr|}
     \right)\, .\label{finwavevec}
\end{eqnarray}   
The last term in the brackets is proportional to the acceleration
of $\gamma _1$, and vanishes if that world-line is freely falling.
Also, the equation (\ref{acceq}) for the time-dilation can be regarded as a consequence of eq. (\ref{finwavevec}), since $d^2\tau _1/d\tau ^2 ={\dot\gamma}_2\cdot\nabla ({\dot\gamma _2}\cdot k/\omega _1)$.

Finally, a physicist receiving signals must decide how to compare successive measurements of $k^a$ along $\gamma _2$.  If $\gamma _2$ is a geodesic, there is a natural choice:  parallel-transport, which leads to the differential formula (\ref{finwavevec}).  If, however, $\gamma _2$ is not a geodesic, one might prefer to use Fermi--Walker transport.  It is easy enough to interconvert the two, the Fermi--Walker derivative of $k^a$ along $\gamma _2$ being \begin{equation}
 {\dot\gamma}_2\cdot \nabla k^a +({\dot\gamma}_2^a{\ddot\gamma}_2^b
-{\ddot\gamma}_2^a{\dot\gamma}_2^b
)k_b\, ,
\end{equation}
so one would supplement (\ref{finwavevec}) by an additional term.  The choice of which quantity to use is really a question of which measures of the change one is most interested in reporting.  The covariant derivative gives us ones less sensitive to the geometry of the world-line $\gamma _2$; the Fermi--Walker derivative is more frame-dependent but measures more directly the changes in frequency and spatial wave-vector relative to the observer.
The same principle will apply to changes in the field.

\subsection{Change in the field}

We may also analyse the effects on the received electromagnetic field of its
propagation through the gravitational field; this includes changes in amplitude and polarisation.  

As is well-known, in geometric optics the field $F_{ab}$ is transverse to the direction $l^a$ of propagation.  That is, however, not a relativistic formulation of the condition; relativistically the statement is that $l^a$ must be a repeated principle null direction of $F_{ab}$ \citep{Pirani1965,PR1986}.
We may express this conveniently by choosing a complex null vector $m^a$ (covariantly constant along $l^a$ and normalised to $l\cdot m=0$, $m\cdot{\overline m}=-1$); then the field must have the form
\begin{equation}\label{phgo}
  F_{ab}=\phi (l_am_b-m_al_b)+\mbox{conjugate}
\end{equation}
for a scalar field $\phi$.  One can fix the freedom in the choice of $m^a$ to have $\phi\geq 0$; then the two-form $l_am_b-m_al_b$ carries the polarisation information.
One has
\begin{equation}
l\cdot\nabla \phi =\rho\phi\, ,
\end{equation}
where $\rho =-(1/2)\nabla\cdot l$ is the convergence of $l^a$. 

It is convenient to introduce a {\em luminosity distance} $\rr$ such that $l\cdot \nabla \rr =-\rho \rr$.  
Then $l\cdot\nabla (\rr F_{ab} )=0$ and $\rr F_{ab}$ is parallel-transported along the null geodesic.  This means that the field $F_{ab}$ diverges as $1/\rr$ at $\gamma _1$, but that is simply a mathematical artefact of modelling the emitter as a point source.  Really, one should imagine a finite surface of emission, and $\rr F_{ab}$ taking direction-dependent limits as one approaches this surface.

The natural normalisation for $\rr$ is then with respect to the world-line of the electromagnetic source, that is
$({\dot\gamma}_1\cdot l)^{-1}l\cdot\nabla \rr\Bigr| _{s_1}=1$.  The direction-dependent limit of 
$\rr F_{ab}$
on $\gamma _1$ is a measure of the intrinsic strength of the field at the source, which will vary along the world-line.

We can express $\rr$ in terms of the quantities already given.  Since $\rho$ measures $-1/2$ the logarithmic rate of increase of the surface area element along the rays abreast $l^a$, the luminosity distance is the square root of the area element.  Taking into account the normalisation, we have
\begin{equation}\label{rreq}
  \rr =({\dot\gamma}_1\cdot l)\sqrt{-2V^a{}_cV^b{}_dm_a{\overline m}_bm^{[c}{\overline m}^{d]}}\, .
\end{equation}

We have then
\begin{eqnarray}
  l\cdot\nabla w\cdot\nabla\left( \rr F_{ab}\right)
  &=&\left[ l\cdot\nabla ,w\cdot\nabla\right] 
    \rr F_{ab}\nonumber\\
  &=&-2l^pw^qR_{pq[a}{}^c\rr F_{b]c}\, .
\end{eqnarray} 
Integrating this, we have
\begin{eqnarray}\label{feelq}
  {\dot\gamma}_2\cdot\nabla\left(\rr F_{ab}\right)
    &=&w\cdot\nabla\left(\rr F_{ab}\right)\Bigr| _{s_2}
           \nonumber\\
    &=&w\cdot\nabla\left(\rr F_{ab}\right)\Bigr| _{s_1}
    \nonumber\\
&&\quad    -2\int _{s_1}^{s_2}l^pw^qR_{pq[a}{}^c\rr F_{b]c}\, d\as \nonumber\\
    &=&\frac{d\tau _1}{d\tau}{\dot\gamma}_1\cdot\nabla\left(\rr F_{ab}\right)\Bigr| _{s_1}\nonumber\\
 &&\quad   -2\int _{s_1}^{s_2}l^pw^qR_{pq[a}{}^c\rr F_{b]c}\, d\as \nonumber\\
 &=&\frac{d\tau _1}{d\tau}{\dot\gamma}_1\cdot\nabla\left(\rr F_{ab}\right)\Bigr| _{s_1}\nonumber\\
 &&\quad   -2\left(\int _{s_1}^{s_2}l^pw^qR_{pq[a}{}^c\, d\as\right)\rr F_{b]c} 
 \, ,\qquad
\end{eqnarray}
where we understand that parallel propagation along the null geodesic has been used to identify $w\cdot \nabla (\rr F_{ab})\Bigr|_{s_1}$
with a tensor at $\gamma _2$, as well as to define the integrals.\footnote{Thus an integral written as $\int _{s_1}^{s_2} Q_{ab}(\as )\, d\as$ is really $\int _{s_1}^{s_2} P_a{}^c(\as )P_b{}^d(\as ) Q_{cd}(\as )\, d\as$, where $P_a{}^c(\as )\lambda _c$ is the result of parallel-propagating $\lambda _c$ to $\gamma _2$ along the null geodesic $p(\as ,\tau )$; similarly we should have $P_a{}^c(s_1)P_b{}^d(s_1)w\cdot\nabla (\rr F_{cd})\Bigr| _{s_1}$ for $w\cdot\nabla (\rr F_{ab})\Bigr| _{s_1}$.}
Then
\begin{eqnarray}\label{fceq}
    {\dot\gamma}_2\cdot\nabla F_{ab}
    &=& -\frac{{\dot\gamma}_2\cdot\nabla\rr}{\rr} F_{ab}
    +\frac{d\tau _1}{d\tau}\frac{{\dot\gamma}_1\cdot\nabla\left(\rr F_{ab}\right)\Bigr| _{s_1}}{\rr}\nonumber\\
    &&-2\left(\int _{s_1}^{s_2}l^pw^qR_{pq[a}{}^c\, d\as\right)\, F_{b]c} \, .\qquad
\end{eqnarray}
The last term is a pleasingly clean Lorentz transformation derived from a curvature integral; the middle term corresponds to changes in the received field due to intrinsic changes in the source.  The first term, due to changes in luminosity distance, can be expressed (somewhat lengthily) in terms of the data we have, as follows.

We have, from eq. (\ref{rreq}),
\begin{eqnarray}\label{logeq}
 {\dot\gamma}_2\cdot\nabla\log\rr
  &=&{\dot\gamma}_2\cdot\nabla\log ({\dot\gamma}_1\cdot l)
  \\
  &&+(1/2){\dot\gamma}_2\cdot\nabla\log
  V^a{}_cV^b{}_dm_a{\overline m}_bm^{[c}{\overline m}^{d]}\, .
  \nonumber
\end{eqnarray} 
Here
\begin{eqnarray}\label{logeqa}
  {\dot\gamma}_2\cdot\nabla {\dot\gamma}_1\cdot l
    &=&w\cdot\nabla {\dot\gamma}_1\cdot l\nonumber\\
    &=&(w\cdot\nabla{\dot\gamma}_1^a)l_a+{\dot\gamma}_1^a
      w\cdot\nabla l_a\nonumber\\
    &=&\frac{d\tau _1}{d\tau}{\ddot\gamma}_1\cdot l
      +{\dot\gamma}_1\cdot\beta\, .
\end{eqnarray}
Differentiating the normalisation conditions for $m^a$, we find
\begin{eqnarray}\label{logeqb}
  {\dot\gamma}_2\cdot\nabla m_{[a}{\overline m}_{b]}
  &=&w\cdot\nabla m_{[a}{\overline m}_{b]}\nonumber\\
  &=&-(l\cdot\nabla U^p{}_q\alpha ^q +l\cdot\nabla V^p{}_q\beta ^q) m_p n_{[a}{\overline m}_{b]}\nonumber\\
  &&-\mbox{ conjugate}\nonumber\\
  &&+\mbox{ terms proportional to } l_a\, ,\ l_b\, ,
\end{eqnarray}
where the terms proportional to $l_a$ or $l_b$ will not contribute to eq. (\ref{logeq}), because those are eigenvectors of $V^p{}_q$.     
Finally, we
must compute ${\dot\gamma}_2\cdot\nabla V^a{}_b=w\cdot\nabla V^a{}_b\Bigr| _{\gamma _2}$.  Differentiating the Jacobi equation (which $V^a{}_b$ satisfies) and working out some commutators, we find
\begin{equation}\label{dweq}
  l\cdot\nabla ^2 w\cdot\nabla V^a{}_b-l^pl^qR_{pcq}{}^a
    w\cdot\nabla V^c{}_b
     =S^a{}_b
\end{equation}
where
\begin{eqnarray}\label{logeqd}
S^a{}_b&=&     (w\cdot\nabla (l^pl^qR_{pcq}{}^a)) V^c{}_b
       -w^pl^qR_{pqc}{}^al\cdot\nabla V^c{}_b\nonumber\\
&&       -l\cdot\nabla(w^pl^qR_{pqc}{}^aV^c{}_b)\, .
\end{eqnarray}
We can regard eq. (\ref{dweq}) as a sort of inhomogeneous Jacobi equation with source $S^a{}_b$, and solve it by variation of parameters.  The result is
\begin{eqnarray}\label{logeqc}
w\cdot\nabla V&=&\int _{s_0}^{s}
\left[ U(s)-V(s)(V^{-1})(\as )U(\as )\right]\times\\
&&\left[l\cdot\nabla U-(l\cdot\nabla V)V^{-1}U\right] ^{-1}(\as )
S(\as )\, d\as\, ,\nonumber
\end{eqnarray}
where the indices have been omitted (and matrix operations are to be understood throughout) in the interest of clarity.

Combining eqs. (\ref{logeqa}), (\ref{logeqb}), (\ref{logeqc}) (evaluated at $s=s_2$) and (\ref{logeqd}) gives the first term on the right in eq. (\ref{fceq}), and this completes the formula for the change in the field at the receiver.

\section{Passage to linearised gravity}

We will now specialise to linearised gravity.  Thus we regard the metric as a first-order perturbation of the Minkowskian one.
We shall nevertheless avoid an explicit choice of gauge, and continue to present the results in an invariant form, in order to keep the geometry and physics as clear as possible.

There is an overall issue to keep in mind in such schemes:  in general, a quantity of interest will have both zeroth- and first-order terms.  If it is not a scalar, and its zeroth-order term is non-vanishing, then a first-order gauge change will in general add in a portion of the zeroth-order term to the first-order term.  This means that the decomposition into zeroth- and first-order terms is not invariant unless the zeroth-order term vanishes.

In what follows, many of the contributions we compute will be purely first-order, and thus will have invariant interpretations.  However, for each of the effects there will also be zeroth-order terms.  (For instance, if $\gamma _1$ and $\gamma _2$ are skew time-like geodesics in Minkowski space, then $\tau _1(\tau )$ will incorporate a time-dependent Doppler effect.)  Thus there will be certain contributions which have no invariant decomposition into zeroth- and first-order effects; for these, any attempt to specify them in terms of a background Minkowski geometry will require choosing a gauge.

The place this mixing of zeroth- and first-order terms will show up is when we use parallel transport along the null geodesic to identify the tangent ${\dot\gamma}_1^a$ to $\gamma _1$ with a vector at $\gamma _2$.  In fact, the quantity which will enter is
\begin{equation}
  \delta ^a = {\dot\gamma}_2^a-\frac{d\tau _1}{d\tau}P^a{}_b{\dot\gamma}_1{}^b\, ,
\end{equation}
where $P^a{}_b$ is the parallel-propagator from $\gamma _1(\tau _1)$ to $\gamma _2(\tau )$ along $p(s,\tau )$.  In the previous section, I did not write $P^a{}_b$ explicitly, but here it is best to do so, to guard against the temptation to use a background Minkowski structure to subtract $(d\tau _1/d\tau ){\dot\gamma}_1^a$ from ${\dot\gamma}_2^a$ (and thus neglect a potential first-order part).  In general, I will not write the parallel-propagator factors in terms which are already first-order (since the omitted corrections would be of higher order), but I will keep them in zeroth-order terms.

For later reference, note that
\begin{eqnarray}\label{deltcheq}
  {\dot\gamma}_2\cdot\nabla \delta ^a
    &=&-{\ddot\gamma}_2^a-\frac{d^2\tau _1}{d\tau ^2}P^a{}_b
       {\dot\gamma}_1{}^b-\frac{d\tau _1}{d\tau} {\dot\gamma}_2\cdot\nabla P^a{}_b{\dot\gamma}_1{}^b\nonumber\\
        &=&{\ddot\gamma}_2^a-\frac{d^2\tau _1}{d\tau ^2}P^a{}_b
       {\dot\gamma}_1{}^b-\frac{d\tau _1}{d\tau} \times\nonumber\\
       &&\left(
       \frac{d\tau _1}{d\tau}P^a{}_b{\ddot\gamma}_1{}^b
       -\int _{s_1}^{s_2} w^pl^qR_{pqb}{}^a{\dot\gamma}_1{}^b\, d\as\right)\, .\qquad
\end{eqnarray}
(In keeping with the remarks above, since the curvature is first-order, I have omitted the parallel-propagator terms which should properly appear in the integrand.)  From this, we have
\begin{eqnarray}\label{deltsqcheq}
\delta _a{\dot\gamma}_2\cdot\nabla\delta ^a
 &=&\frac{d\tau _1}{d\tau} {\ddot\gamma}_{2a}P^a{}_b{\dot\gamma}_1^b -\left(\frac{d\tau _1}{d\tau}\right)^2 {\dot\gamma}_{2a}P^a{}_b{\ddot\gamma}_1^b\nonumber\\
 && -\frac{d^2\tau _1}{d\tau ^2}\left( {\dot\gamma}_{2a}P^a{}_b{\dot\gamma}_1^b -\frac{d\tau _1}{d\tau}\right)\nonumber\\
&&  +\frac{d\tau _1}{d\tau}\int _{s_1}^{s_2} R_{pqab}w^pl^q{\dot\gamma}_1^a{\dot\gamma}_2^b\, d\as\, .
\end{eqnarray}  
In these equations, and in others that follow, the field $w^a$ appears within terms which are already first-order.  In such terms, we need only the zeroth-order expression for $w^a$; this is
\begin{equation}\label{veeq}
  v^a(\as )=
 \frac{s_2-\as}{s_2-s_1} \frac{d\tau _1}{d\tau}{\dot\gamma}_1^a
   +\frac{\as -s_1}{s_2-s_1}{\dot\gamma}_2^a
     \, ,
\end{equation} 
which linearly interpolates between the values of $w^a$ at the ends.
(Because this will be used only when multiplied by first-order factors, we do not write the parallel propagators which transport the vectors at the ends of the null geodesic to $p(\as ,\tau )$.)

The remaining quantities we shall need are the solutions $U^a{}_b$, $V^a{}_b$ to the Jacobi equation, the vector $\beta ^a$ which is one of the initial data (the vector $\alpha ^a=(d\tau _1 /d\tau ){\dot\gamma}_1^a$ is already known), and the luminosity distance $\rr$.

To first order in the metric perturbation, we have
\begin{eqnarray}
 U^a{}_b&=&\left(\delta ^a{}_c+u^a{}_c\right) P^c{}_b\\
 V^a{}_b&=&\left((s-s_1)\delta ^a{}_c +v^a{}_c\right) P^c{}_b\,
\end{eqnarray}
where now $P^c{}_b=P^c{}_b(s)$ is the parallel-propagator along $p(s,\tau )$ from $\gamma _1(\tau _1(\tau ))=p(s_1,\tau )$ to $p(s,\tau )$ and
\begin{eqnarray} 
 u^a{}_b&=&\int _{s_1}^{s_2} (s-\as )H(s-\as )l^pl^qR_{pbq}{}^ad\as\label{ueq}\\
  v^a{}_b&=&\int _{s_1}^{s_2} (s-\as )(\as -s_1) H(s-\as )l^pl^qR_{pbq}{}^ad\as\, ,\quad\label{veq}
\end{eqnarray}  
with $H$ the Heaviside step-function, and the curvature is evaluated at $p(\as ,\tau)$ in the integrands.  

We find, in the linear approximation, that
\begin{eqnarray}
  \beta ^a &=&\frac{1}{s_2-s_1}(P^{-1})^a{}_b\delta ^b
   -\frac{d\tau _1/d\tau}{s_2-s_1}u^a{}_b{\dot\gamma}^b_1
  \nonumber\\  
&&  -\frac{1}{(s_2-s_1)^{2}}v^a{}_b\delta ^b\, ,
\end{eqnarray}
where on the right the first term contains the zeroth-order contribution, and $u^a{}_b$, $v^a{}_b$ are evaluated at $s=s_2$.

Finally, from eqs. (\ref{veq}) and (\ref{rreq}), we find
\begin{eqnarray}\label{lumdis}
\rr &=&{\dot\gamma}_1\cdot l\left( (s_2-s_1) -v^a{}_b{\overline m}_am^b\right)\nonumber\\
&=&{\dot\gamma}_1\cdot l\left( \phantom{\int _{s_1}^{s_2}}(s_2-s_1)\right.\\
&&\left. -4\pi G\int _{s_1}^{s_2} (s_2-\as )(\as -s_1) l^pl^qT_{pq}\, d\as\right)\
\, ,\nonumber
\end{eqnarray}
where $T_{ab}$ is the stress--energy and we have used Einstein's equation.

\subsection{Basic results and discussion}

With the results of the beginning of this section and a bit
of work using eqs.~(\ref{alpheq}), (\ref{beteq}), (\ref{acceq}), we have
\begin{eqnarray}\label{timeq}
\frac{d^2\tau _1}{d\tau ^2}&=&
  (l\cdot{\dot\gamma}_1)^{-1}
    \left[ (s_2-s_1)^{-1}\delta\cdot\delta
      +\int _{s_1}^{s_2} l^pl^q R_{paqb}
   v^av^b\, d\as\right]\nonumber\\
    &&+(l\cdot{\dot\gamma}_1)^{-1} l\cdot{\ddot\gamma}_2
    -(l\cdot{\dot\gamma}_1)^{-3}(l\cdot{\dot\gamma}_2)^2l\cdot{\ddot\gamma}_2
   \, .
\end{eqnarray}

A similar computation, using eq.~(\ref{finwavevec}), gives us the rate of change of the wave-vector
$k^a=\omega _1l^a$.  
We have
\begin{eqnarray}\label{wavechange}
{\dot\gamma}_2\cdot\nabla  k^a
  &=&\frac{\omega _1}{l\cdot{\dot\gamma}_1}\frac{\delta ^a}{s_2-s_1}
    +\frac{\omega _1}{l\cdot{\do\gamma}_1}
     \int _{s_1}^{s_2}l^pl^q v^b\frac{\as -s_1}{s_2-s_1} R_{pbq}{}^a\, d\as\nonumber\\
     &&-\frac{k^a}{l\cdot{\dot\gamma}_1}
     \left[ l\cdot{\dot\gamma}_2l\cdot{\ddot\gamma}_1
       +\frac{{\dot\gamma}_{1c}(P^{-1})^c{}_b\delta ^b}{s_2-s_1}
       \phantom{\int _{s_1}^{s_2}}\right.\nonumber\\
      &&\left.   -{\dot\gamma}_{1c}\int _{s_1}^{s_2}l^pl^q v^b\frac{s_2-\as}{s_2-s_1} R_{pbq}{}^c\, d\as\right]\, .
\end{eqnarray}    

And for the change in field, we find
\begin{eqnarray}\label{fch}
{\dot\gamma}_2\cdot\nabla F_{ab}
&=&\frac{s_2-s_1}{\rr}\left( \frac{d\tau _1}{d\tau} {\ddot\gamma}_1\cdot l +\frac{{\dot\gamma}_1\cdot \delta}{s_2-s_1}
  \right.\nonumber\\
&&\left.  -\int _{s_1}^{s_2} l^pv^cl^q{\dot\gamma}_1^d \frac{s_2-\as}{s_2-s_1} R_{pcqd}\, d\as\right) F_{ab}\nonumber\\
&&-\frac{4\pi G}{\rr}\times\nonumber\\
&&\left(
  \int _{s_1}^{s_2} (s_2-\as)(\as-s_1)v\cdot\nabla l^pl^qT_{pq}\, d\as\right) F_{ab}\nonumber\\
&&  +\frac{d\tau _1}{d\tau}\frac{{\dot\gamma}_1\cdot\nabla (\rr F_{ab})\Bigr| _{s_1}}{\rr}\nonumber\\
  &&-2\left(\int _{s_1}^{s_2} l^pv^qR_{pq[a}{}^c\, d\as\right)
    F_{b]c}\, .     
\end{eqnarray}

Equations (\ref{timeq}), (\ref{wavechange}) and (\ref{fch}) describe basic observables, in the limit of linearised gravity.  There are several points to make about these results:

(a) The possibility of ascribing an observation of one the left-hand quantities to gravitational radiation relies on having some sort of extra information allowing one to discriminate between the various terms on the right.  In most cases, we must assume that time-dependence of the gravitational waves is enough different from those of the other physical processes that this can be used.

(b) The equations contain terms proportional to $\delta ^a$ or $\delta\cdot\delta$ divided by $l\cdot {\dot\gamma}_1 (s_2-s_1)$.  These terms typically are suppressed when the observer is at very great distances from the emitter.  (In fact, in view of the formulas
(\ref{deltcheq}), (\ref{deltsqcheq}), (\ref{lumdis}), for monochromatic waves of angular frequency $\omega$, the first-order parts of these terms are typically suppressed by factors of $(\omega \rr )^{-1}$.)  In such cases, the zeroth-order contribution to $d^2\tau _1/d\tau ^2$ may become effectively negligible, leaving a geometrically pure first-order curvature-integral term.

(c) In vacuum, the second term on the right-hand side of eq. (\ref{fch}) will vanish.  Note that, apart from this term,
all of the curvature-integral terms are sums of
\begin{equation}\label{Ieeq}
I^{(n)}_{ab}=\int _{s_1}^{s_2}\left(\frac{\as}{s_2-s_1}\right) ^n l^pl^qR_{paqb}\, d\as
\end{equation}
for $n=0,1,2$, where we have used the formula (\ref{veeq}) for $v^a$.
(Note that the form (\ref{phgo}) of $F_{ab}$ implies this for (\ref{fch}).)

(d) Many traditional approaches to these problems aim to work out what we might call the long-range scattering, corresponding to the receiver and emitter receding to great distances along the same null geodesic.\footnote{It might be tempting to call this the total scattering, but that would be misleading in this context, because we still track here only the differential effects, as the light-rays vary, of the scattering.}  The formulas here show, however, that only in restricted circumstances will this limit exist.

For the long-range scattering, we want to examine what happens as
$s_1\to -\infty$, $s_2\to +\infty$.  Of course, we must assume that the contributions from the accelerations ${\ddot\gamma}_1^a$, ${\ddot\gamma}_2^a$ are negligible (or at least are stable under the limit), and that the $\delta ^a$-dependent terms drop out.  But even then the remaining curvature integrals will {\em not} in general stabilise.  
This is because they depend on the quantity $v^a(\as )$ (eq. (\ref{veeq})) which interpolates from $(d\tau _1/d\tau ){\dot\gamma}_1^a$ at $\gamma _1$ to ${\dot\gamma}_2^a$ at $\gamma _2$.  
This quantity has no well-defined limit point-wise in $\as$ as $s_1\to -\infty$, $s_2\to +\infty$ independently.  In other words, the contributions of the curvature integrals are in principle sensitive to the choices of of $s_1$ and $s_2$ in the asymptotic regime.

On the other hand, in many cases of interest ${\dot\gamma}_1^a$ and ${\dot\gamma}_2^a$ will differ by only sub-relativistic effects, and then $v^a(\as )$ will be nearly constant along the null geodesic.  Then the curvature integrals contributing to the time-dilation(\ref{timeq}) and field change (\ref{fch}) will (assuming the curvature falls off suitably) stabilise as $s_1\to -\infty$, $s_2\to +\infty$.

Below, we shall mostly be interested in the case where the receiver is removed to arbitrarily great distances, but the emitter is held fixed.  In this case we will have $v^a\to (d\tau _1/d\tau ){\dot\gamma}_1^a$, and only the integrals (\ref{Ieeq}) for $n=0$ will contribute.

(e) Besides the implicit dependence of $v^a(\as )$ on $s_1$, $s_2$, the curvature integrals in the expression (\ref{wavechange}) for the change in wave-vector involve explicit factors  $(\as-s_1)/(s_2-s_1)$, $(s_2-\as )/(s_2-s_1)$.  These factors are of geometric origin, and express the fact that change in angle perceived by a distant observer will be of order half the full scattering angle multiplied by the ratio $($distance of source to scatterer$)/($distance of source to receiver$)$.  In practice this means that if the source of the light is much closer to the source of the gravitational waves than it is to the Earth, the angular change due to the scattering is correspondingly reduced.

Thus attempts to measure angular deflections due to gravitational waves are at a geometric disadvantage relative to measurements of changes in time-dilation or field.
(This point can also be deduced from the formulas in \citealt{KSGU}.)

(f) Below, we shall be interested in the case where the receiver is very distant but the emitter is not, so $s_2\to +\infty$ but $s_1$ is finite.  In this case we will have $v^a\to (d\tau _1/d\tau ){\dot\gamma}_1^a$, and the change in wave-vector will be suppressed.

\section{Quadrupole Sources}

With the formulas derived above, the analysis of the effects at the linearised level in any given space--time reduces to the computation of certain moments of the curvature over the relevant segment of the light-ray's trajectory.  The curvature can itself be expressed as a retarded field due to sources, plus a possible pure radiation term.  
The results of this can be quite complicated, even in simple cases, because of the time-dependence of the curvature and the different components which enter, and the fact that we wish to take the point of emission of the light to be finite.  
However, for the remainder of this paper, the aim will not be detailed modelling but simply rough estimates of the scales of the effects.

I shall here work out the leading contributions in a simple but important case:  a pure quadrupole field, from an isolated source, with the null geodesics in the radiation zone in the sense that $\omega b\gg 1$, where $\omega$ is the angular frequency of any component of the gravitational wave and $b$ is the null geodesic's impact parameter relative to the quadrupole source.  

The point of reception will be taken to be very far away from the source; this corresponds to the limit $s_2\to +\infty$ discussed earlier, but $s_1$ will be held finite (recall $s_1$, $s_2$ are the affine parameters specifying the null geodesic segment from emission to reception).  In this case, the only curvature integral (\ref{Ieeq}) we have to compute is $I^{(0)}_{bd}$ (because within the integrands we have
$v^a\to (d\tau _1/d\tau ){\dot\gamma}_1^a$, $((s_2-\as)/(s_2-s_1))\to 1$, $((\as -s_1)/(s_2-s_1))\to 0$).

Of course, real sources have monopole and perhaps dipole as well as quadrupole components; however, these contribute only stationary terms to the field, and in any event in the linear approximation those can simply be added to the quadrupole effects.

\subsection{Quadrupole fields}

By a quadrupole field I mean a linearised gravitational field in an appropriate $j=2$ representation of the rotation group; both `electric' and `magnetic' quadrupoles (often called mass quadrupole and current quadrupole terms) are allowed.  The treatment here is chosen to fit with rest of this paper's formalism; other forms are given in \citet{RW1957,Pirani1965,Thorne1980}.

Let us first consider the `electric' part, which we idealise as a pure quadrupole at the spatial origin.  (Since we are only interested in the field outside the source, this is adequate.)  Let the quadrupole moment be $Q_{ab}^{\rm el}(t)$, with arbitrary time-dependence.  (Here $t$ is the the coordinate time at the spatial origin, and $Q_{ab}^{\rm el}$ is symmetric, trace-free and orthogonal to $t^a$.)  It makes a contribution
\begin{equation}\label{QT}
 T_{ab}^{\rm el} =(1/2)(t\cdot\nabla \delta ^p{}_a-t_a\nabla ^p)
   (t\cdot\nabla \delta ^q{}_b-t_b\nabla ^q)Q_{pq}^{\rm el}(t)\delta ^{(3)}
\end{equation}
to the stress--energy, where $\delta ^{(3)}$ is the spatial delta-function.  One easily verifies that $\int (t^pt^qT_{pq}^{\rm el}) (x_a -tt_a)(x_b-tt_b) \, d^3 x =Q^{\rm el}_{ab}$ -- the mass quadrupole is indeed $Q^{\rm el}_{ab}$.  

The full curvature tensor can easily be worked out by standard means.\footnote{One point to be careful of in these calculations is that $f(u)\nabla _a\delta ^{(3)}\not= f(t)\nabla _a\delta ^{(3)}$ in general, as becomes clear by multiplying by a test function and integrating by parts.}  It is convenient to introduce a null tetrad $l^a$, $m^a$, ${\overline m}^a$, $n^a$, with $l^a=t^a+{\hat r}^a$, for ${\hat r}^a$ a unit space-like radially outward vector, $n^a=t^a-{\hat r}^a$, and $m^a=2^{-1/2}(\partial _\theta -i\csc\theta \partial _\phi )$.  Then the radiative (order $r^{-1}$) term is 
\begin{equation}
R_{abcd}^{\rm el}=-4G \frac{Q^{{\rm el}\,(4)}_{pq}{\overline m}^p{\overline m}^q}{r} l_{[a}m_{b]}l_{[c}m_{d]}+\,\mbox{conjugate}\,
+
\cdots\, ,
\end{equation}
where the superscript $(4)$ indicates the order of differentiation with respect to $u$.  (Note that the polarisation factors $l_{[a}m_{b]}$ are the same as in the electromagnetic case.)
One does not actually need the detailed form of the $m^a$ vectors in computations; 
the combination ${\overline m}_pm_b$ entering here may be written as
\begin{equation}
  {\overline m}_p m_b=(1/2)(-g_{pb}+t_pt_b-{\hat r}_p{\hat r}_b +i\epsilon _{pbqs}t^q{\hat r}^s )\, .
\end{equation}

Weyl curvatures of `magnetic' type can be obtained, in linearised theory in the vacuum, by dualizing the electric ones.  Thus a `magnetic' quadrupole field will, in the vacuum region, be given by
\begin{equation}
R_{abcd}^{\rm mag}=4iG \frac{Q^{{\rm mag}\,(4)}_{pq}{\overline m}^p{\overline m}^q}{r} l_{[a}m_{b]}l_{[c}m_{d]}+\,\mbox{conjugate}\,
+
\cdots\, ,
\end{equation}
where $Q^{\rm mag}_{pq}$ is referred to as the magnetic part of the quadrupole moment.\footnote{The sign here is fixed by the convention that it is the real and imaginary parts of the Bondi shear which determine the electric and magnetic parts of the curvature.}
The form of the source for this term is different from (\ref{QT});
one can check that the corresponding contribution to the stress--energy is
\begin{equation}
T_{ab}^{\rm mag} =(1/2)t^p\epsilon _{pqr(a}\left(-{\dot Q}_{b)}^{{\rm mag}\, q}+t_{b)}Q_s^{{\rm mag}\, q}\nabla^s\right)\nabla ^r\delta ^{(3)}\, .
\end{equation}
It turns out that $Q^{\rm mag}_{ab}$ is essentially a first spatial moment of the angular momentum density.  To see this, note that ${\mathcal L}^a=-\epsilon _{pqr}{}^a(t^p)(x^q)(t^cT_c{}^r)$ can be interpreted as the angular momentum density with respect to the spatial origin (the minus sign giving the usual convention for the angular momentum as a spatial vector).  Then a short calculation shows
\begin{equation}
  \int {\mathcal L}_a(x_b-tt_b)\, d^3 x=\frac{3}{4} Q^{\rm mag}_{ab}\, .
\end{equation}

One may take a complex quadrupole moment $Q_{ab}=Q^{\rm el}_{ab}+iQ^{\rm mag}_{ab}$; then the curvature in the radiation zone is
\begin{equation}\label{curv}
R_{abcd}=-4G \frac{{\overline Q}^{(4)}_{pq}{\overline m}^p{\overline m}^q}{r} l_{[a}m_{b]}l_{[c}m_{d]}+\mbox{conjugate}
+
\cdots\, .
\end{equation}

\subsection{The light-rays}

We wish to study the differential scattering of light-rays which pass through the gravitational wave source's radiation zone and are received at some great distance.  In this subsection, we work out the appropriate parametrization of those rays.  Since we are to compute an integral of the curvature, which is a first-order quantity, it is enough to know the geometry of the ray and of the receiver to zeroth order.

Let the Bondi coordinates, centred at the world-line of the gravitational-wave source, be $(u,r,\theta ,\phi )$.  Actually, we will not need to write $(\theta ,\phi )$ explicitly; we may represent them by their corresponding null vector $\al ^a=\al ^a(\theta ,\phi )$, normalised by $\al \cdot t=1$, with $t^a$ the unit future-directed time-like vector characterising the Bondi frame.  A point in Minkowski space is thus specified as $ut^a+r\al ^a$.  We will suppose the light-ray is received at an event
\begin{equation}
  \gamma _2^a(\tau )=u_2t^a+r_2l_2^a
\end{equation}
where $r_2$ is very large.

In general, the equation of a light-ray may be expressed conveniently as
\begin{equation}
 p^a(s)=(u-b)t^a +bl_0^a+sl_1^a\, ,
\end{equation}
where $u$ is the retarded time the light-ray tends to as $s\to +\infty$, the null future-directed vectors $l_0^a$, $l_1^a$ are normalised by $l_0\cdot l_1=l_0\cdot t=l_1\cdot t=1$ (so the spatial parts of $l_0^a$, $l_1^a$ are orthogonal); then $b$ is the ray's impact parameter, its point of closest approach to the spatial origin occurs at $s=0$, and $l_0^a$, $l_1^a$ code the direction of closest approach and the direction of the ray.  We have $r=\sqrt{b^2+s^2}$ and $\al ^a =(b/r)l_0^a+(s/r)l_1^a +(1-b/r-s/r)t^a$.  

\begin{figure}
\includegraphics[width=3.25in]{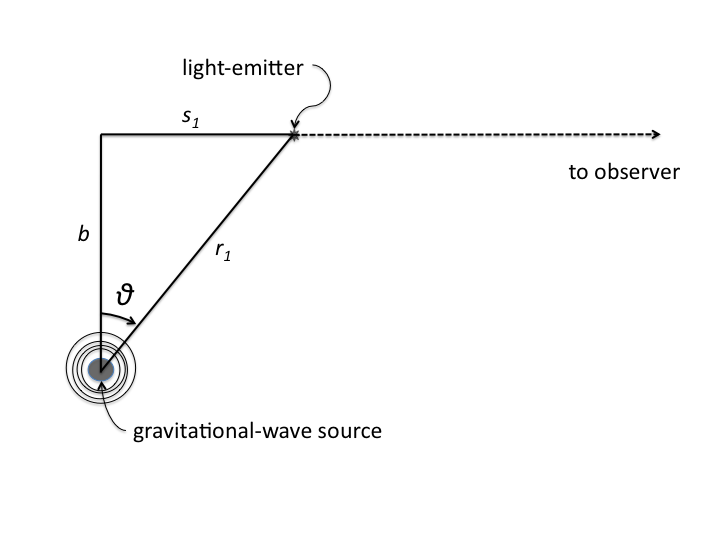}
\caption{\label{fig:Geom} The geometry of the gravitational-wave source and the light-emitter.  The impact parameter is $b$.  The coordinate along the light-ray is $s$, with $s=0$ corresponding to the ray's closest approach to the gravitational-wave source, and $s_1$ the value at which the light-emitter sits; the observer is at a very large value $s_2$.  The distance between the gravitational-wave source and a point on the light-ray is $r$, with $r_1$ the distance of the light-emitter.  Then angle $\theta$ (not used until eq. (\ref{thdef})) is given by $s_1=r_1\sin\theta$; note that $s_1$ and $\theta$ may have either sign.}
\end{figure}

Requiring the ray to be received at $p^a(s_2)=\gamma ^a_2(\tau )$, we find
\begin{eqnarray}
  u+s_2&=&u_2+r_2\\
  b^2+s_2^2&=&r_2^2\\
  (b^2+s_2^2)^{-1/2} (bl_0^a+s_2l_1^a)
    &=&l_2^a\mbox{  mod  } t^a\, ,
\end{eqnarray}
where `mod $t^a$' means up to terms proportional to $t^a$.
Solving these equations perturbatively in $b/r_2$, we have
\begin{eqnarray}
  s_2&=&r_2+O(b^2/r_2)\\
  u&=&  u_2+O(b^2/r_2)\\
  l_1^a&=&l_2^a+O(b/r)\, .
\end{eqnarray}
To this order, the vector $l^a_0$ is (apart from the normalisations specified above) unrestricted; the quantity $(b/r_2)l_0^a$ specifies the apparent direction of the light-ray, relative to the direction of the source, at the receiver.    

The light-ray's trajectory is thus
\begin{equation}
 p^a(s)=(u_2-b)t^a+bl_0^a +sl_2^a\, ,
\end{equation}
meeting the receiver at $s=s_2=r_2$.  The point of emission will be at a parameter value $s_1$, the point of closest approach to the spatial origin, as noted above, would be $s=0$.
Expressing $p^a(s)$ in Bondi coordinates, we have
\begin{equation}
 p^a(s)=ut^a +r\al ^a\, ,
\end{equation} 
where
\begin{eqnarray}
  u&=&u_2+s-\sqrt{b^2+s^2}\label{uueq}\\
  r&=&\sqrt{b^2+s^2}\\
  \al ^a&=& (b/r)l_0^a +(s/r)l_1^a +(1-b/r-s/r)t^a\, .
\end{eqnarray}  

In the computations to follow, it will be convenient to put
\begin{equation}
  s=b\sinh \xi \, .
\end{equation}
Then $r=b\cosh\xi$ and $bs-\sqrt{b^2+s^2}=-be^{-\xi}$ and we define
\begin{equation}\label{Seq}
 S=-e^{-\xi} =(s/b)-\sqrt{1+(s/b)^2}\, ;
\end{equation}
note that $S$ is an increasing function of $s$, and that
\begin{eqnarray}
  ds&=&b\cosh\xi\, d\xi \nonumber\\
  &=&r\, d\xi\nonumber\\
  &=&-r\frac{dS}{S}\, .
\end{eqnarray}

\subsection{The curvature integral}

As noted above, the only curvature integral we require is
\begin{equation}
  I^{(0)}_{bd}=\int _{s_1}^{\infty}l^al^c R_{abcd}\, d\as\, ,
\end{equation}
where $l^a=l^a_1$.  The integrand, in the radiation zone, is
\begin{eqnarray}
l^al^cR_{abcd} &=&-4G\frac{{\overline Q}^{(4)}_{pq}}{r}
  l^a\al _{[a}\am_{b]}\oam^p
  l^c\al _{[c}\am_{d]}\oam^q\nonumber\\
 &&+\,\mbox{conjugate}\,+\cdots\, ,
\end{eqnarray}  
where the accents indicate the vectors evaluated at $\as$ along the null geodesic.  

Because we wish to evaluate $I^{(0)}_{bd}$ in the case $\omega b\gg 1$, where $\omega$ is the angular frequency of any contributing component to the gravitational radiation field, we shall for the moment just work with one Fourier component, putting
$e^{i\omega u}K_{pq}$
(where $K_{pq}$ is constant, symmetric, trace-free, and orthogonal to $t^a$)
in place of $Q_{pq}$; after using the condition $\omega b\gg 1$ we will restore $Q_{pq}$.  Then
making use of eqs. (\ref{uueq}) and (\ref{Seq}),  we find
\begin{eqnarray}
I^{(0)}_{bd}&=&4G\omega ^4\int _{S_1}^0 e^{-i\omega (u_2+bS)}{\overline K}_{pq}
  l^a\al _{[a}\am_{b]}\oam^p
  l^c\al _{[c}\am_{d]}\oam^q\, \frac{dS}{S}\nonumber\\
  &&+\,\mbox{conjugate}\,+\cdots\, .
\end{eqnarray}  
In this form, the integral is proportional to $e^{- i\omega bS}$, and we may regard it as effecting a Fourier transform.  We are interested in the behaviour of this for large $\omega b$, which is to say the high-frequency regime.  The function to be transformed is smooth except for being cut off at the end-points $S=S_1$, $S=0$; it is the non-smooth behaviour at these end-points which will give the leading contribution.

In fact, it is the lower end-point $S=S_1$ which makes the dominant contribution, for a little algebra shows the integrand tends continuously to zero as $S\uparrow 0$.  We have then the Fourier transform of a function with a jump discontinuity at $S=S_1$; this is
\begin{eqnarray}\label{Ieee}
I^{(0)}_{bd}&\sim &
-4iG\omega ^3{\overline K}_{pq}\times\nonumber\\
&&\int _{S_1}^0 e^{-i\omega (u_2+bS)}  l^a\al _{[a}\am_{b]}\oam^p
  l^c\al _{[c}\am_{c]}\oam^q \frac{dS}{S}\Bigr| _{S=S_1}\nonumber\\
  &&+\,\mbox{conjugate}\,+\cdots
 \nonumber\\
  &\sim&-4G {\overline Q}^{(3)}_{pq}l^a\al _{[a}\am_{b]}\oam^p
  l^c\al _{[c}\am_{d]}\oam^q (bS)^{-1}\Bigr| _{S=S_1}\nonumber\\
  &&+\,\mbox{conjugate}\,+\cdots\, ,
\end{eqnarray}
where the tilde denotes asymptotic expansion for $\omega b\gg 1$ and we have restored $Q_{pq}$.

We will be most interested in the case of time-delays, for which the curvature-integral contribution is
\begin{equation}
\frac{d^2\tau _1}{d\tau ^2} =(d\tau _1 /d\tau )^2I^{(0)}_{bd}t^bt^d\, .
\end{equation}
Then the combination of vectors entering into eq. (\ref{Ieee}) becomes
\begin{eqnarray}
4t^bl^a\al _{[a}\am _{b]}\oam ^p
&=&-2l^a\am _a\oam ^p\Bigr| _{\as =s_1}\\
&=&\left[\frac{b}{\ar}\left( \frac{b}{\ar}l_1^p-\frac{\as}{\ar} l_0^p\right)
 -i\frac{b}{\ar}C^p\right]\Bigr| _{\as =s_1} 
 \mbox{ mod } t^p\, ,\nonumber
\end{eqnarray}
where terms proportional to $t^p$ have been dropped because they will not contribute when contracted with $Q_{pq}$ and
\begin{equation}
  C^p=\epsilon ^p{}_{aqs}l_1^at^ql_0^s
\end{equation}
is a purely spatial vector equal to ${\bf l}_0\times {\bf l}_1$, where ${\bf l}_0$, ${\bf l}_1$ are the spatial parts of $l_0^a$, $l_1^a$.  Writing now $r_1=\ar\Bigr| _{\as=s_1}=\sqrt{b^2 +(s_1)^2}$, we have
\begin{multline}
\frac {d^2\tau _1}{d\tau ^2} \sim \left(\frac{d\tau _1}{d\tau}\right) ^2\frac{Gb^2}{4(r_1-s_1)(r_1)^2}\\
\times {\overline Q}^{(3)}_{pq}
\left(\frac{b}{r_1}l_1^p-\frac{s_1}{r_1} l_0^p-i\frac{b}{r_1}C^p\right)
\left(\frac{b}{r_1}l_1^q-\frac{s_1}{r_1} l_0^q-i\frac{b}{r_1}C^q\right) \\
 +\,\mbox{conjugate}\,+\cdots\, ,\mskip100mu
\end{multline}
or
\begin{multline}\label{ttimeq}
\frac {d^2\tau _1}{d\tau ^2} 
  \sim \left(\frac{d\tau _1}{d\tau}\right) ^2\frac{Gb^2}{(r_1-s_1)(r_1)^2}\\
\times  \left\{ \frac{1}{2} Q^{{\rm el}\, (3)}_{pq}\left[
  \left(\frac{b}{r_1} l_1^p-\frac{s_1}{r_1}l_0^p\right)
  \left(\frac{b}{r_1} l_1^q-\frac{s_1}{r_1}l_0^q\right)
  -\frac{b^2}{(r_1)^2}C^pC^q\right] \right.\\ 
 \left.
   -Q^{{\rm mag}\, (3)}_{pq}
  \left(\frac{b}{r_1} l_1^p-\frac{s_1}{r_1}l_0^p\right)
  \left(\frac{b}{r_1} C^q\right)\right\} 
   +\cdots\, .
\end{multline}  
Here $Q^{(3)}_{pq}$ is evaluated at $u=u_2+s_1-r_1$.

We may recast the foregoing in
terms of the angle $\theta$ of the light-emission from the point of closest approach relative to the source (see Fig.~\ref{fig:Geom}), so
\begin{equation}\label{thdef}
  s_1=r_1\sin\theta\, ,\quad b=r_1\cos\theta
\end{equation}
(where $b$ is the impact parameter);
then we have  
\begin{multline}\label{ttimeqtheta}
\frac {d^2\tau _1}{d\tau ^2} 
  \sim \left(\frac{d\tau _1}{d\tau}\right) ^2\frac{G (1+\sin\theta)}{r_1}\\
\times  \left\{ \frac{1}{2} Q^{{\rm el}\, (3)}_{pq}\left[
  \left( l_1^p\cos\theta -l_0^p\sin\theta\right)
  \left( l_1^q\cos\theta-l_0^q\sin\theta\right)
  -C^pC^q\cos ^2\theta\right] \right.\\ 
 \left.
   -Q^{{\rm mag}\, (3)}_{pq}
  \left( l_1^p\cos\theta -l_0^p\sin\theta\right)
  \left( C^q\cos\theta\right)\right\} 
   +\cdots\, .
\end{multline}  
Formula 
(\ref{ttimeqtheta}) is the main result, relating the clock times $\tau _1$ of the emitter (at radius $r_1$) and $\tau$ of the receiver, as influenced by gravitational mass (`electric') $Q^{\rm el}_{ab}$ and current (`magnetic') $Q^{\rm mag}_{ab}$ quadrupole sources (evaluated at retarded time $u=u_2+s_1-r_1=u_2-(1-\sin\theta ) r_1$, where $u_2$ is the observer's retarded time).
Recall that here $l_1^a$ is the null geodesic's tangent and $l_0^a$ is the null vector whose spatial part ${\bf l}_0$ is a unit vector from the source to the geodesic's point of closest approach, and $C^a$ is a spatial vector ${\bf l}_0\times {\bf l}_1$ normal to the plane containing the source and the light-ray.

Perhaps the most striking feature of this result is the `forward-backward' asymmetry represented by the overall factor $(1+\sin\theta)$, which enhances effects from light-emitters on the portion of the light-ray outgoing from the gravitational-wave source ($0<\theta <\pi /2$) relative to those from the incoming portion ($-\pi /2<\theta <0$).  This is a relativistic effect arising from the use of light-signals to probe the space--time curvature.  For gravitational waves of a given frequency with respect to $t^a$ (the gravitational source's frame), the frequency with respect to an affine parameter along the light-ray will be larger along the ingoing portion than along the outgoing one, and the effects due to those higher-frequency terms more nearly average out.

For the scaling of eq. (\ref{ttimeqtheta}) with distance, for light-emitters very distant from the 
light-ray's point of closest approach to the
gravitational-wave source, that is $r_1\gg b$, we have $1+\sin\theta =1\pm\sqrt{1-(b/r_1)^2}$ and 
\begin{equation}
\frac{1+\sin\theta}{r_1} =\begin{cases}
  b^2/(2r_1^3)&\mbox{for }\theta\downarrow -\pi /2\\
  2/r_1&\mbox{for }\theta\uparrow \pi /2\, .
  \end{cases}
\end{equation}  
Thus the rough magnitude of the effect will be
\begin{equation}\label{rough}
\left| \frac {d^2\tau _1}{d\tau ^2}\right| \sim
\begin{cases}
\left| d\tau _1/d\tau\right| ^2  
  \left\| {Q}^{(3)} _{pq}\right\| (Gb^2/(r_1^3c^4))
   &\mbox{for }\theta\downarrow -\pi /2   \\
  \left| d\tau _1/d\tau\right| ^2  
    \left\| {Q}^{(3)} _{pq}\right\| (G/(bc^4))
&\mbox{for moderate }\theta
\vphantom{\Biggr|}%%To prevent the double bars joining
   \\
\left| d\tau _1/d\tau\right| ^2 
  \left\| {Q}^{(3)} _{pq}\right\| (G/(r_1c^4)) 
&\mbox{for }\theta\uparrow\pi /2\, ,
   \end{cases}
\end{equation}  
where the speed of light has been given explicitly.
(For rough estimates, the precise choice of norm for the tensor is not very important; any $L_p$ norm in terms of a standard Euclidean basis will do.)
While the upper line corresponds to the scaling found by Damour and Esposito-Far\`ese, we see that the fall-off for light-sources along the outgoing portion of the ray is much softer, having the $\sim r^{-1}$ behaviour characteristic of radiative effects.

While the appearance of this radiative scaling is certainly of interest, we shall see below that even in favourable circumstances the effects are small; we shall therefore concentrate, in the following sections, with the case of moderate $\theta$, corresponding to light-emitters with $r_1\sim b$, the middle line of eq. (\ref{rough}).

Finally, two remarks about the angular dependence of the effects through the factor in curly braces in eq. (\ref{ttimeqtheta}).
First, the variations of this term for moderate $\theta$ means that different light-emitters in this regime will probe the different components of the quadrupole tensors.
Second,
one might have thought that, for light-sources further away, the trigonometric factors would tend to suppress the dependences on $l_0^a$, the vector from the origin to the point of closest approach, and lead to a dependence of the effects primarily on $l_1^a$, the tangent to the light-ray.  The opposite is true of the factor in curly braces in eq. (\ref{ttimeqtheta}), however.  
This is a direct consequence of the transversality of the waves; the components of the curvature that enter are orthogonal to the position-vector relative to the origin.

\section{Estimates}

Of the various possible modulations of the light by its passage through, and emission within, the gravitational radiation -- 
changes in the received light's amplitude, polarisation, phase, location on the receiver's sky, and time-dilation -- it seems that in most cases time-dilation will be the most promising for detection (but still, as we shall see, quite challenging).  As pointed out earlier, changes in the location and phase will be suppressed by a geometric factor.  Changes in amplitude and polarisation would be so small they would probably be too hard to detect.

In this section I shall estimate two time-dilation effects, red-shift and pulse time-of-arrival offsets, in some cases of interest.  First some rough general formulas for these will be derived; we will see that they are expressed naturally in terms of a dimensionless intrinsic measure of gravitational-wave strength, the {\em Bondi news} (essentially $GQ_{ab}^{(3)}/c^5$ in our case).  While this quantity is central in much of gravitational radiation theory, few numerical values for it have appeared in the astrophysical literature, so samples of these are given.  

I then discuss the pulsar timing resolutions appropriate to the detection of time-of-arrival wanderings (unfortunately, the extraordinary long-term stability of pulsars does not help directly with this), and finally estimate those wanderings.

\subsection{Formulas for the red-shift and time-offsets}

The entire light-signal from the emitter is subject to a time-dilation, which will itself be a function of time.  If the emitted signal is $a(\tau _1)$ in the frame of the emitter, then, the received signal will be (neglecting other effects) $a(\tau _1(\tau ))$; that is, there will be a distortion due to the relative differences in the flows of time.  One effect this would give rise to would be a time-dependent red-shift.    Or if one knew the light-source was, in its own frame, emitting regular signals (for instance, if it were a pulsar), the effect of the gravitational radiation would be to make the times of receipt of these wander slightly from complete regularity.  

Where linearised gravity is adequate, the equation governing this was (\ref{timeq}), and it contained three sorts of contributions:   terms due to possible covariant accelerations of the emitter or the receiver; a sort of kinematic term due to the possible boost of the receiver relative to the emitter; and the curvature integral estimated in the previous section.  It is the last which will be important, as will now be explained.

Recall that our main hope for detecting gravitational waves by this effect comes not from relative magnitudes of these terms (the gravitational acceleration at the surface of the Earth is far larger than the expected gravitational-radiation effects), but from the different time-dependences of the terms.  We must assume that we can account for any source and receiver accelerations well enough to distinguish the effects from those of possible gravitational waves.  

The kinematic term $\delta \cdot \delta /(l\cdot {\dot\gamma}_1(s_2-s_1))$ requires a bit more discussion, though.  Here $\delta ^a$ is the difference between $(d\tau _1 /d\tau ){\dot\gamma}_1^a$ and ${\dot\gamma}_2^a$, parallel-propagated along the null geodesic.  
While in the situations we shall consider the zeroth-order contribution to this will have a different time-dependence than the gravitational-wave effects, the first-order contribution will be affected by the gravitational radiation. 
However the corresponding effects are suppressed by a factor of $(\omega r/c)^{-1}$ (essentially because they involve velocities rather than accelerations), as inspection of eqs. (\ref{deltsqcheq}), (\ref{wavechange}) shows; see the discussion under point (b) in 3.1.  

We thus consider only the curvature-integral contributions to the time-dilation.
Assuming that the light-source is not moving ultra-relativistically with respect to the Earth, we see from eq. (\ref{rough}) that the scale of the effects is, in the case of light-sources $\sim b$ from the gravitational-wave source, where $b$ is the light-ray's impact parameter,
\begin{equation}\label{ruff}
\left| \frac {d^2\tau _1}{d\tau ^2}\right| \sim
\frac{c}{b}\cdot\left( \frac{G
  }{c^5}\left\| {Q}^{(3)} _{pq}\right\|\right)\, .
\end{equation}  
where the first factor on the right has units inverse time and the second is dimensionless (recall $Q^{(3)}_{ab}$ is the third time-derivative of the quadrupole moment).\footnote{The right-hand side of eq. (\ref{ruff}) also gives a rough estimate of the fractional rate of change of the electromagnetic field components, induced by the light's passage through, and emission in, the gravitational radiation.}

The second factor on the right in eq. (\ref{ruff}) is a dimensionless measure of the intrinsic strength of the gravitational radiation; it is essentially an average value 
of the magnitude of the {\em Bondi news} $N$, a key quantity in gravitational radiation theory which will be discussed a bit further below.  Because this is such central concept, we now switch to expressing the quantities of interest in terms of this average news 
\begin{equation}\label{mndef}
\mnews =\frac{G
  }{c^5}\left\| {Q}^{(3)} _{pq}\right\|\, .
\end{equation}
Then the rough magnitudes of the red-shifts will be
\begin{equation}
| z|\sim  \frac{c}{\omega b}\cdot \mnews 
\end{equation}
and the magnitude the time-of-arrival wandering will be
\begin{equation}
  |\Delta\tau |\sim \frac{c}{\omega ^2b}\cdot \mnews \,
\end{equation}
where $\omega$ is the angular frequency of the wave and $b$ is impact parameter of the null geodesic from the gravitational wave source.  (Again we assume the emission occurs near the point of closest approach -- more generally one would replace the factors of $1/b$ with $(1+\sin\theta )/r_1$, in accord with (\ref{ttimeqtheta}) --, and that the motion of the receiver relative to the emitter is sub-relativistic.)

We recall that the analysis of the previous section assumed that the light emission occurred in the gravitational wave zone.  This means that we must have $\omega b /c\gtrsim 1$.  (Emissions from points closer to the gravitational source could be studied by the general formulas given earlier, but they would correspond to near- or intermediate-zone gravitational disturbances, which had not propagated far enough for their wave character to be fully developed.)
Thus we have
\begin{equation}
| z|_{\rm max}\sim   \mnews 
\end{equation}
and 
\begin{equation}\label{taumax}
  |\Delta\tau |_{\rm max}\sim \omega ^{-1} \mnews 
\end{equation}
for rough estimates of the largest possible red-shift and time-offset due to light emissions near the geodesic's point of closest approach, in the gravitational wave zone, for a given gravitational-wave source.

\begin{table*}
 \centering
 \begin{minipage}{140mm}
  \caption{Rough estimates of the Bondi news $\mnews$ for some classes of gravitational-wave sources, inferred from estimates of the gravitational luminosity (and, for the last four classes, the quadrupole approximation).  The first three classes are burst sources.}
  \begin{tabular}{@{}llllc@{}}
  \hline
   Source type&Estimate&Parameters and expected ranges\\
      \hline
 highly asymmetric supernovae&possibly a significant fraction of unity&\\
  mergers of comparable-mass black holes&possibly a significant fraction of unity&\\
 neutron-star glitches\footnote{Adapted from \citet{AFJ2011}.}&  $8\times 10^{-5} \left(\frac{E_{\rm gw}}{10^{-6}M_\odot c^2}\right) ^{1/2}\left(\frac{1\, {\rm ms}}{\tau _{\rm decay}}\right)^{1/2}$ &$E_{\rm gw}=$ energy in the wave\\
  &&$\tau _{\rm decay}=$ decay time\\
 &&$10^{-9}M_\odot <E_{\rm gw}<10^{-6}M_\odot$\\
 long-duration waves from neutron stars\footnote{Adapted from \citet{KAA2001}, \citet{Prix2009}.}%
   &$3\times 10^{-6} \varepsilon\left(\frac{M}{1.4M_\odot}\right) 
  \left(\frac{R}{10\, {\rm km}}\right)^2
  \left(\frac{10\, {\rm ms}}{P}\right) ^3$&$\varepsilon=$ efficiency\\
  &&$\varepsilon\lesssim 2\times 10^{-5}$ for periods as long as years\\
  binary, equal masses $M$, circular orbit%
  &$2\times 10^{-16}\left(\frac{M}{M_\odot}\cdot\frac{1\, {\rm d}}{P}\right)^{5/3}$&$P=$ period\\
  binary, masses $M\gg m$, circular orbit%
  &$1\times 10^{-14}\left(\frac{m}{M_\odot}\right)
      \left(\frac{M}{100M_\odot}\right) ^{2/3}
      \left( \frac{1\, {\rm d}}{P}\right)^{5/3}$&$P=$ period\\
 \hline
\end{tabular}
\end{minipage}
\end{table*}

\subsection{Source types and intrinsic wave strengths}

In much of the literature, gravitational radiation is estimated by a combination $h$ of the linearised metric components at the Earth.  While useful for discussions involving terrestrial detectors, this measure is neither invariant nor intrinsic; it is not well-suited for the present considerations.

The measure which is appropriate is
the {\em Bondi news} $\news$, which we have already mentioned.  This is a function of retarded time and angle; 
it has a suitable invariance (Bondi--Metzner--Sachs covariance), is intrinsic and dimensionless \citep{PR1986}.
The gravitational luminosity is $(c^5/4\pi G)\oint |\news |^2$ (over the sphere of directions).
In the quadrupole approximation one has $\news =-(G/c^5){\overline Q}^{(3)}_{ab}{\overline m}^a{\overline m}^b$ (where ${\overline m}^a$ is a complex null vector coding the direction), and so the quantity $\mnews$ given in eq. (\ref{mndef}) is an average of $|\news|$ over directions.
There are, however, few numerical values of the news in the literature.

Table 1 gives some rough estimates for the news, using $\mnews =( GL/c^5)^{1/2}$ where $L$ is the gravitational luminosity (and, for the last four lines, the quadrupole approximation).
Note that the first three cases correspond to burst-type sources.

It is clear from Table 1 that if we should be lucky enough to observe light-signals from the vicinity of an extreme gravitational-wave event, such as an asymmetric supernova or colliding black holes of comparable mass, a great deal of information could be gained.  However, because the gravitational waves decay very quickly in such cases (typically on a time-scale of order the light-crossing time associated with the mass, that is $GM/c^3$), one would have to already have the telescope trained on the object, and in general this would require extraordinary serendipity.  An exception would be if we could detect an inspiralling system and thus be prepared to monitor light-sources in its vicinity when strong gravitational waves were produced.

For the rest of this paper, we leave aside the possibility of
gravitational-wave sources with news any significant fraction of unity.  The best other candidates for detection appear to have binaries as sources of gravitational waves, corresponding to the last two lines in Table 1.  For these, we have
\begin{equation}\label{eqmalg}
 \mnews\Bigr| _{m=M} \sim 1.3\times \left(\frac{GM}{c^3}\cdot\frac{2\pi}{P}\right) ^{5/3}
\end{equation} 
for the case of equal masses in a circular orbit and
\begin{equation}\label{uneqmalg}
  \mnews \Bigr| _{M\gg m}\sim  2.5\times \left(\frac{2\pi Gm}{c^3P}\right)
    \left( \frac{2\pi GM}{c^3P}\right)^{2/3}
\end{equation} 
for highly unequal masses $M\gg m$ in a circular orbit; in each case $P$ is the period.

\subsection{Red-shifts}

For neutron-star glitches, the factor $8\times 10^{-5}$ in the third line of Table~1 corresponds to a spectral resolving power of $20\, {\rm km}\, {\rm s}^{-1}$.  
The red-shift will be smaller by a factor $(c/\omega b)$, and for a light-source ten light-seconds away from the glitching neutron star (say, a companion) this is likely $\sim 10^{-4}$ (taking $1 /\omega\sim 1$ ms), giving $|z|\sim 2\, {\rm m\, s}{}^{-1}/c$.  Radial components of stellar velocities are measured, interferometrically, to $1-3\, {\rm m\, s}{}^{-1}$ \citep{Butler1996,Rupprecht2004}; however, such precisions typically require integration times of a minute or more, whereas the glitches are thought to decay over milliseconds.

For the other sources we considered, the red-shifts would be smaller, and hence we now move on to the possibilities of detecting the wanderings of times-of-arrival of signals from pulsars.

\subsection{Pulse time-of-arrival offsets}

We now come to the possibility of measuring the offsets caused in pulsar signals times-of-arrival caused by their passage through gravitational waves.  We saw above
(eq. (\ref{taumax})) that the maximum offset was roughly $\omega ^{-1}\mnews$.  The effect is therefore determined by a ratio of the intrinsic gravitational wave strength $\mnews$ to the gravitational wave period $\omega$, and we must consider this for different classes of sources.

First, however, let us consider what temporal resolution we might get from the pulsar signals.

\subsubsection{Millisecond pulsars as clocks}

As account of pulsar and timing is given by 
\citet{LorimerKramer2005}.
While millisecond pulsars can have extraordinary long-term stability, that stability is essentially a secular effect, based on measuring the times and numbers of pulses between initial and final pulses years apart, and having only to reckon with the uncertainties in locating the times of arrival of those initial and final signals.  Except in searching for very low frequency gravitational waves (or `memory' effects, which are essentially zero-frequency phenomena), that stability is not of direct help.  We must consider timing over shorter scales.

There are two key issues:  that pulsars are weak radio sources, and that their signals over a period appear to be subject to  intrinsic variations, jitter and drifting sub-pulses.  Because of these, one cannot extract very accurate times of arrival from individual pulses; instead one must fold them, that is, integrate (typically over a few minutes) with a mean periodicity removed in order to get a good pulse profile.  One may then have a resolution around $\Delta\tau _{\rm pulsar}\sim 10^{-7}-10^{-6}\, {\rm s}$.

If we search for effects corresponding to shifts in the times of arrival of order $\Delta \tau \ll \Delta \tau _{\rm pulsar}$, we need of order $(\Delta \tau _{\rm pulsar}/\Delta \tau )^2$ samplings, each of some minutes' length.  There are about $5\times 10^5$ minutes in a year, and so it would be very hard to track changes smaller than $\sim 10^{-10}\, {\rm s}$ by this crude statistical means with current technology.

However, were we to learn more about the jitter and sub-pulse drift, we might be able to do better.  Most of the effects considered here will have fairly well-defined periods, and the question is whether those Fourier components could be distinguished from the jitter and drift.
These displacements of the times-of-arrival would show up in the Fourier domain as splittings of the angular frequencies of the components of the signal by $\pm\omega$, or as peaks in the Fourier-transformed residuals, at angular frequencies $\pm\omega$.

\subsubsection{Long-duration waves from neutron stars}

The case where the pulsar is one member of a binary, the other member being a neutron star gravitationally radiating due to an asymmetry or instability (fourth line of Table 1), turns out not to be very promising.
(Note that here the gravitational radiation in question is {\em not} due to the binary system.)  Taking the case of a non-axisymmetric neutron star considered by \citet{Prix2009}, we have
\begin{eqnarray}
  \Delta\tau &\sim& \mnews (c/\omega ^2b)\nonumber\\
&\sim&\left(6\times 10^{-6}\, {\rm s}\right)|\news |\left(\frac{P}{10\, {\rm ms}}\right)^2
  \left(\frac{10\,\mbox{lt-s}}{b}\right)\, ,
\end{eqnarray}
where $P$ is the period of rotation.
Using Table 1, we see that
in optimistic circumstances for persistent waves, we might have effects as large as $\sim 4\times 10^{-16}\, {\rm s}$.  This is very small.  

\subsubsection{Pulsar detection of binary emission}

\begin{figure}
\includegraphics[width=3.25in]{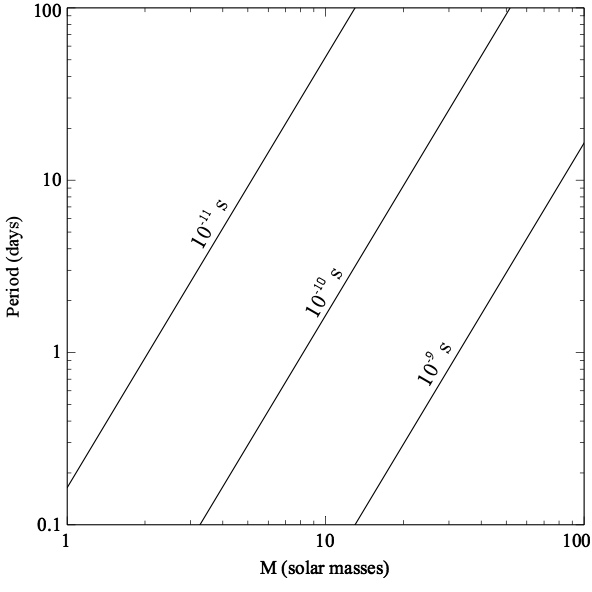}
\caption{\label{fig:emf} Contours of approximate maximum time-offsets $\Delta\tau _{\rm max}$ from two equal masses, each $M$, in a circular orbit of period $P$.}
\end{figure}

It is more promising to consider as a source of gravitational waves a tight binary, and assume that there is a nearby pulsar.  This pulsar could be a (relatively distant) tertiary, or not be directly gravitationally bound but still close by (as might happen if, for example, both the binary and the pulsar were in a globular cluster).  
Recalling that
$\omega b/c\gtrsim 1$ 
for the wave character of the binary-induced changes in curvature to be developed at the pulsar,
and using eqs. (\ref{eqmalg}), (\ref{taumax}), we see that
for a binary of equal masses, each $M$, in a circular orbit, the maximum offset in pulse time-of-arrival will be
\begin{equation}\label{emc}
\Delta \tau _{\rm max}\Bigr| _{M=m}\sim\left(  3\times 10^{-12}\,{\rm s}\right)\left(\frac{M}{M_\odot}\right)^{5/3}\left(\frac{1\, {\rm d}}{P}\right)^{2/3}
  \, .
\end{equation}

Figure \ref{fig:emf} shows contours for $\Delta\tau _{\rm max}$ for (\ref{emc}).
For example, if we had the component masses $M=5M_\odot$ and period $P=.1\, {\rm d}$,  we would have
$\Delta \tau _{\rm max}\sim 2\times 10^{-10}\, {\rm s}$.  This (rough) estimate of the maximum time-displacement corresponds to the pulsar being at the inner part of the gravitational radiation zone, that is
$\sim cP/(2\pi)\sim 30 \, {\mbox{l-min}}$ away from the  binary.  If the pulsar were as far away as $10^{-2}\, {\rm pc}$ (representative of interstellar distances in a globular cluster), one would have $\Delta\tau \sim 3\times 10^{-13}\, {\rm s}$.

\begin{figure}
\includegraphics[width=3.25in]{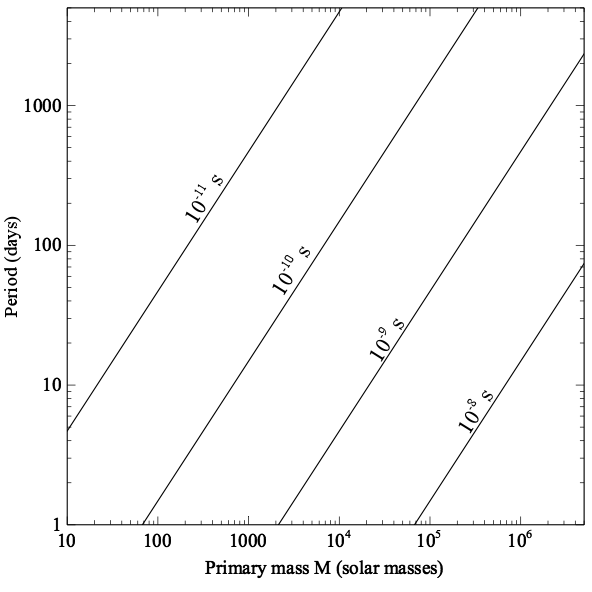}
\caption{\label{fig:umf} Contours of approximate maximum time-offsets $\Delta\tau _{\rm max}$ from two unequal masses, with primary mass $M\gg m$ and secondary mass $m=M_\odot$, in a circular orbit of period $P$.}
\end{figure}

In the case $M\gg m$ of one mass dominating the other we have (from eqs. (\ref{uneqmalg}), (\ref{taumax}))
\begin{equation}
\Delta \tau _{\rm max}\Bigr| _{M\gg m}
\sim \left(1\times 10^{-10}\, {\rm s}\right)
\left(\frac{m}{M_\odot}\right)
      \left(\frac{M}{100M_\odot}\cdot \frac{1\, {\rm d}}{P}\right)^{2/3}\, .
\end{equation}
Contours of $\Delta\tau _{\rm max}$ for this case, for the secondary of mass $m=M_\odot$, are given in Fig. \ref{fig:umf}.  
For instance, if one could find a black hole of mass $M=10^4M_\odot$ (in, say, a globular cluster), with a star with mass $m=M_\odot$ orbiting in a one-day period, one would have $\Delta \tau _{\max}\sim 3\times 10^{-9}\, {\rm s}$; again, this limit is set when the pulsar is at the inner part of the gravitational wave zone, in this case a few light-days away from the radiating binary.
For a solar-mass star in a ten-year circular orbit about the $M=4\times 10^6M_\odot$ presumed black hole at the galactic centre, we should have $\Delta\tau _{\rm max}\sim 7\times 10^{-10}\, {\rm s}$, and this would apply to pulsars around $cP/(2\pi )\sim 1.6$ l-yr from that system.

It should be emphasized that these estimates are very rough.

\section{Summary and Discussion}

The first main goal of this paper was to present an invariant framework for treating the differential scattering of light-rays in exact general relativity.  This allows one to keep track of the different physical contributions to the various scattering effects, and to focus on quantities of direct physical interest.

The result of greatest near-term possible observational consequence is that light emitted from the vicinities of gravitational-wave sources may be scattered by much larger amounts than those discussed by Damour and Esposito-Far\`ese, and by Kopeikin, Sch\"afer, Gwinn and Eubanks.  The results here are however compatible with those authors'; one can understand them as due to `edge effects', whose possibility was explicitly noted by Damour and Esposito-Far\`see, and which could be treated within the general formalism of Kopeikin et al.  An equivalent statement is that for light emissions at finite distances from the gravitational-wave sources, the scattering depends on the gravitational radiation field, in contrast to the cancellations found by Damour, Kopeikin et al. which occur in the limit of infinite distances.
The effects do come out to be, roughly, of the scale of those predicted in some of the still earlier papers \citep{Sazhin1978,Fak1994}, and in a very rough sense one may say that this is because those papers did have the scattering respond to the gravitational radiation fields; however (again in accord with Damour, Kopeikin et al.) the details of the arguments of those papers are not really compatible with those here.

The main physical issue which was not addressed systematically in earlier work was the effect of the gravitational waves on the motion of the light-sources.  (See however \citealt{KEK2011} for an exception.)  This showed up, mathematically, in leaving the results in terms of the coordinate times rather than clock times of the light-emitter and receiver.  Such formulas are of direct physical significance only if the coordinates are adapted to the motions of both the light-emitter and receiver.  This issue is particularly problematic when the light-emitter is close enough to the gravitational-wave source that that gravitational effect on the light-emitter's motion must be accounted for.  The present approach works with the clock times and can account for any motions of the light-emitter and receiver.

Some candidates for the possible observation of these scattering effects were considered, in a quadrupole approximation for the gravitational waves.  It was found that the best effects to search for were based on the relative distortion of the emitter's proper time $\tau _1$ as measured by light-signals incoming to the receiver at its proper time $\tau$.  

Because these observations would require training a telescope on the emitter, one is driven to look for specific likely sources of gravitational radiation, and these are in general weak ones.  The most promising sort identified here would be a tight binary, the light-emitter being a pulsar orbiting the binary as a distant tertiary, or not gravitationally bound but close by.  In an example with favourable but not extreme parameters (a binary of two $5M_\odot$ stars with a $.1\, {\rm d}$ period, and a pulsar about $30$ l-min away), we found offsets in the times of arrival of the pulsar pulses of the order of $\sim 10^{-10}\, {\rm s}$, with the offsets varying with a period around an hour.  If we a were able to find a solar-mass star orbiting a $10^4M_\odot$ black hole with a one-day period, for example in a globular cluster, pulsars a few light-days away might have their pulse times-of-arrival offset by $\sim 10^{-9}\, {\rm s}$.

\section*{Acknowledgments}

I thank Sergei Kopeikin and Bahram Mashhoon for drawing my attention to earlier work on this and for a helpful discussion.
I thank Duncan Lorimer for a helpful electronic exchange about pulsar timing measurements, and an anonymous referee for constructive suggestions about the exposition.
The graphs were prepared with \textsc{Veusz}.  This work was supported in part by the University of Missouri Research Board.

%\bibliographystyle{../Library/texmf/tex/latex/mnras/bibtex/bst/mnras/mn2e}

%\bibliographystyle{mn2e}

%\bibliography{references}

%\begin{thebibliography}{99}

\label{lastpage}

\end{document}